\documentstyle[epsf,epsfig,here,12pt]{article}  
\begin{document}
\newcommand{\bea}{\begin{eqnarray}}
\newcommand{\eea}{\end{eqnarray}}
\newcommand{\HR}{\rule{1em}{0.4pt}}
\newcommand{\rhobar}{\bar {\rho}}
\newcommand{\etabar}{\bar{\eta}}
\newcommand{\epsilonk}{\left|\varepsilon_K \right|}
\newcommand{\vubovcb}{\left | \frac{V_{ub}}{V_{cb}} \right |}
\newcommand{\vubsvcb}{\left | V_{ub}/V_{cb}  \right |}
\newcommand{\vtdovts}{\left | \frac{V_{td}}{V_{ts}} \right |}
\newcommand{\epsp}{\frac{\varepsilon^{'}}{\varepsilon}}
\newcommand{\dmd}{\Delta m_d}
\newcommand{\dms}{\Delta m_s}
\newcommand{\pr}{{\rm P.R.}}
\newcommand{\Ds}{{D}_s^+}
\newcommand{\Dp}{{D}^+}
\newcommand{\Do}{{D}^0}
\newcommand{\piss}{\pi^{\ast \ast}}
\newcommand{\pis}{\pi^{\ast}}
\newcommand{\bbar}{\bar{b}}
\newcommand{\cbar}{\bar{c}}
\newcommand{\Dstar}{{D}^{\ast}}
\newcommand{\Dstars}{{D}^{\ast +}_s}
\newcommand{\Dstaro}{{D}^{\ast 0}}
\newcommand{\Dstarstar}{{D}^{\ast \ast}}
\newcommand{\Dbar}{\bar{{D}}}
\newcommand{\Bbar}{\bar{{B}}}
\newcommand{\Bsbar}{  \bar {B}^0_s  }
\newcommand{\Lcbar}{\bar{\Lambda^+_c}}
\newcommand{\nubar}{{\bar \nu_{\ell}}}
\newcommand{\tautaubar}{\tau \bar{\tau}}
\newcommand{\Vcb}{\left | { V}_{cb} \right |}
\newcommand{\Vub}{\left | { V}_{ub} \right |}
\newcommand{\Vud}{\left | {V}_{ud} \right |}
\newcommand{\Vus}{\left | { V}_{us} \right |}
\newcommand{\Vtd}{\left | {V}_{td} \right |}
\newcommand{\Vts}{\left | { V}_{ts} \right |}
\newcommand{\fleisher}{\frac{BR({ B}^0~(\bar{{ B}}^0) 
\rightarrow \pi^{\pm} {\rm K}^{\mp})}
{BR({\rm B}^{\pm} \rightarrow \pi^{\pm} {\rm K}^0)}}
\newcommand{\bptre}{\rm b^{+}_{3}}
\newcommand{\bp}{\rm b^{+}_{1}}
\newcommand{\bo}{\rm b^0}
\newcommand{\bos}{\rm b^0_s}
\newcommand{\bss}{\rm b^s_s}
\newcommand{\qq}{\rm q \bar{q}}
\newcommand{\cc}{\rm c \bar{c}}
\newcommand{\BsDmX}{{B_{s}^{0}} \rightarrow D \mu X}
\newcommand{\BsDsm}{{B_{s}^{0}} \rightarrow D_{s} \mu X}
\newcommand{\BsDsX}{{B_{s}^{0}} \rightarrow D_{s} X}
\newcommand{\BDsX}{B \rightarrow D_{s} X}
\newcommand{\BDomX}{B \rightarrow D^{0} \mu X}
\newcommand{\BDpmX}{B \rightarrow D^{+} \mu X}
\newcommand{\Dsfmn}{D_{s} \rightarrow \phi \mu \nu}
\newcommand{\Dsfipi}{D_{s} \rightarrow \phi \pi}
\newcommand{\DsfX}{D_{s} \rightarrow \phi X}
\newcommand{\DpfX}{D^{+} \rightarrow \phi X}
\newcommand{\DofX}{D^{0} \rightarrow \phi X}
\newcommand{\DfX}{D \rightarrow \phi X}
\newcommand{\DsD}{B \rightarrow D_{s} D}
\newcommand{\DsmX}{D_{s} \rightarrow \mu X}
\newcommand{\DmX}{D \rightarrow \mu X}
\newcommand{\Zbb}{Z^{0} \rightarrow \rm b \bar{b}}
\newcommand{\Zcc}{Z^{0} \rightarrow \rm c \bar{c}}
\newcommand{\Rbb}{\frac{\Gamma_{Z^0 \rightarrow \rm b \bar{b}}}
{\Gamma_{Z^0 \rightarrow Hadrons}}}
\newcommand{\Rcc}{\frac{\Gamma_{Z^0 \rightarrow \rm c \bar{c}}}
{\Gamma_{Z^0 \rightarrow Hadrons}}}
\newcommand{\bb}{\rm b \bar{b}}
\newcommand{\str}{\rm s \bar{s}}
\newcommand{\Bs}{ B^0_s }
\newcommand{\Bsb}{ \bar {B}^0_s }
\newcommand{\Bp}{{B^{+}}}
\newcommand{\Bm}{{B^{-}}}
\newcommand{\Bo}{{B^{0}}}
\newcommand{\Bd}{ {B}^{0}_{d} }
\newcommand{\Bdb}{ \bar {B}^{0}_{d}  }
\newcommand{\Lb}{\Lambda^0_b}
\newcommand{\Lbb}{\bar{\Lambda^0_b}}
\newcommand{\Kstar}{\rm{K^{\star 0}}}
\newcommand{\phim}{\rm{\phi}}
\newcommand{\Dsp}{\mbox{D}_s^+}
\newcommand{\Dn}{\mbox{D}^0}
\newcommand{\Dsb}{\bar{\mbox{D}_s}}
\newcommand{\Dm}{\mbox{D}^-}
\newcommand{\Dnb}{\bar{\mbox{D}^0}}
\newcommand{\Lc}{\Lambda_c}
\newcommand{\Lcb}{\bar{\Lambda_c}}
\newcommand{\Dstarp}{{D}^{\ast +}}
\newcommand{\Dstarm}{{D}^{\ast -}}
\newcommand{\Dsstarp}{\mbox{D}_s^{\ast +}}
\newcommand{\Km}{\mbox{K}^-}
\newcommand{\Pb}{P_{b-baryon}}
\newcommand{\KKpi}{\rm{ K K \pi }}
\newcommand{\GeV}{\rm{GeV}}
\newcommand{\MeV}{\rm MeV}
\newcommand{\nb}{\rm{nb}}
\newcommand{\Zzero}{{\rm Z}^0}
\newcommand{\MZ}{\rm{M_Z}}
\newcommand{\MW}{\rm{M_W}}
\newcommand{\GF}{\rm{G_F}}
\newcommand{\Gm}{\rm{G_{\mu}}}
\newcommand{\MH}{\rm{M_H}}
\newcommand{\MT}{\rm{m_{top}}}
\newcommand{\GZ}{\Gamma_{\rm Z}}
\newcommand{\Afb}{\rm{A_{FB}}}
\newcommand{\Afbs}{\rm{A_{FB}^{s}}}
\newcommand{\sigmaf}{\sigma_{\rm{F}}}
\newcommand{\sigmab}{\sigma_{\rm{B}}}
\newcommand{\NF}{\rm{N_{F}}}
\newcommand{\NB}{\rm{N_{B}}}
\newcommand{\Nnu}{\rm{N_{\nu}}}
\newcommand{\RZ}{\rm{R_Z}}
\newcommand{\fbdsqbd}{f_{B_d} \sqrt{\hat B_{B_d}}}
\newcommand{\fbssqbs}{f_{B_s} \sqrt{\hat B_{B_s}}}
\newcommand{\BK}{B_K}
\newcommand{\rhob}{\rho_{eff}}
\newcommand{\Gammanz}{\rm{\Gamma_{Z}^{new}}}
\newcommand{\Gammani}{\rm{\Gamma_{inv}^{new}}}
\newcommand{\Gammasz}{\rm{\Gamma_{Z}^{SM}}}
\newcommand{\Gammasi}{\rm{\Gamma_{inv}^{SM}}}
\newcommand{\Gammaxz}{\rm{\Gamma_{Z}^{exp}}}
\newcommand{\Gammaxi}{\rm{\Gamma_{inv}^{exp}}}
\newcommand{\rhoZ}{\rho_{\rm Z}}
\newcommand{\thw}{\theta_{\rm W}}
\newcommand{\swsq}{\sin^2\!\thw}
\newcommand{\swsqmsb}{\sin^2\!\theta_{\rm W}^{\overline{\rm MS}}}
\newcommand{\swsqbar}{\sin^2\!\bar{\theta}_{\rm W}}
\newcommand{\cwsqbar}{\cos^2\!\bar{\theta}_{\rm W}}
\newcommand{\swsqb}{\sin^2\!\theta^{eff}_{\rm W}}
\newcommand{\eepm}{{e^+e^-}}
\newcommand{\eeX}{{e^+e^-X}}
\newcommand{\gaga}{{\gamma\gamma}}
\newcommand{\mumu}{\ifmmode {\mu^+\mu^-} \else ${\mu^+\mu^-} $ \fi}
\newcommand{\eeg}{{e^+e^-\gamma}}
\newcommand{\qqb}{{q\bar{q}}}
\newcommand{\Lamp}{\Lambda_{+}}
\newcommand{\Lamm}{\Lambda_{-}}
\newcommand{\Pt}{\rm P_{t}}
\newcommand{\Rb}{\mbox{R}_b}
\newcommand{\Rc}{\mbox{R}_c}
\newcommand{\al}{a_l}
\newcommand{\vl}{v_l}
\newcommand{\af}{a_f}
\newcommand{\vf}{v_f}
\newcommand{\ael}{a_e}
\newcommand{\ve}{v_e}
\newcommand{\amu}{a_\mu}
\newcommand{\vmu}{v_\mu}
\newcommand{\atau}{a_\tau}
\newcommand{\vtau}{v_\tau}
\newcommand{\ahatl}{\hat{a}_l}
\newcommand{\vhatl}{\hat{v}_l}
\newcommand{\ahate}{\hat{a}_e}
\newcommand{\vhate}{\hat{v}_e}
\newcommand{\ahatmu}{\hat{a}_\mu}
\newcommand{\vhatmu}{\hat{v}_\mu}
\newcommand{\ahattau}{\hat{a}_\tau}
\newcommand{\vhattau}{\hat{v}_\tau}
\newcommand{\vtildel}{\tilde{\rm v}_l}
\newcommand{\avsq}{\ahatl^2\vhatl^2}
\newcommand{\Ahatl}{\hat{A}_l}
\newcommand{\Vhatl}{\hat{V}_l}
\newcommand{\Afer}{A_f}
\newcommand{\Ael}{A_e}
\newcommand{\Aferb}{\bar{A_f}}
\newcommand{\Aelb}{\bar{A_e}}
\newcommand{\AVsq}{\Ahatl^2\Vhatl^2}
\newcommand{\Iwk}{I_{3l}}
\newcommand{\Qch}{|Q_{l}|}
\newcommand{\roots}{\sqrt{s}}
\newcommand{\pT}{p_{\rm T}}
\newcommand{\mt}{m_t}
\newcommand{\Rechi}{{\rm Re} \left\{ \chi (s) \right\}}
\newcommand{\up}{^}
\newcommand{\abscosthe}{|cos\theta|}
\newcommand{\dsum}{\Sigma |d_\circ|}
\newcommand{\zsum}{\Sigma z_\circ}
\newcommand{\sint}{\mbox{$\sin\theta$}}
\newcommand{\cost}{\mbox{$\cos\theta$}}
\newcommand{\mcost}{|\cos\theta|}
\newcommand{\epair}{\mbox{$e^{+}e^{-}$}}
\newcommand{\mupair}{\mbox{$\mu^{+}\mu^{-}$}}
\newcommand{\taupair}{\mbox{$\tau^{+}\tau^{-}$}}
\newcommand{\gamgam}{\mbox{$e^{+}e^{-}\rightarrow e^{+}e^{-}\mu^{+}\mu^{-}$}}
\newcommand{\fullskip}{\vskip 16cm}
\newcommand{\halfskip}{\vskip  8cm}
\newcommand{\quarskip}{\vskip  6cm}
\newcommand{\abitskip}{\vskip 0.5cm}
\newcommand{\ba}{\begin{array}}
\newcommand{\ea}{\end{array}}
\newcommand{\bc}{\begin{center}}
\newcommand{\ec}{\end{center}}
\newcommand{\be}{\begin{equation}}
\newcommand{\ee}{\end{equation}}
\newcommand{\eeq}{\end{eqnarray}}
\newcommand{\bes}{\begin{eqnarray*}}
\newcommand{\ees}{\end{eqnarray*}}
\newcommand{\Kz}{\ifmmode {\rm K^0_s} \else ${\rm K^0_s} $ \fi}
\newcommand{\Zz}{\ifmmode {\rm Z^0} \else ${\rm Z^0 } $ \fi}
\newcommand{\qqbar}{\ifmmode {\rm q\bar{q}} \else ${\rm q\bar{q}} $ \fi}
\newcommand{\ccbar}{\ifmmode {\rm c\bar{c}} \else ${\rm c\bar{c}} $ \fi}
\newcommand{\bbbar}{\ifmmode {\rm b\bar{b}} \else ${\rm b\bar{b}} $ \fi}
\newcommand{\xxbar}{\ifmmode {\rm x\bar{x}} \else ${\rm x\bar{x}} $ \fi}
\newcommand{\rphi}{\ifmmode {\rm R\phi} \else ${\rm R\phi} $ \fi}
\newcommand{\bt}{\begin{tabular}}
\newcommand{\et}{\end{tabular}}
\newcommand{\msbar}{\overline{\rm MS}}
\newcommand{\ri}{\rm RI-MOM}
\newcommand{\mtms}{m_t^{\scriptsize{\overline{\rm
MS}}}(m_t^{\scriptsize{\overline{\rm MS}}})}
\newcommand{\msms}{m_s^{\scriptsize{\overline{\rm
MS}}}(\mu=2\, \mbox{GeV})}

\renewcommand{\arraystretch}{1.2}
\pagestyle{empty}
\pagenumbering{arabic}
\vspace*{-1.8cm}
\begin{flushright}
{LAL 00-77} \\
{ROME1-1307/00} \\
{RM3-TH/00-16} \\
%{GE/{\bf MANCA !!!}....} 
\end{flushright}
\vskip  1.5 cm
\begin{center}
{\LARGE {\bf  2000 CKM-TRIANGLE ANALYSIS}}
\vskip .4 cm 
{\Large {\bf  A Critical Review with Updated Experimental
}}\vskip .2 cm 
{\Large {\bf  Inputs and  Theoretical Parameters}}
\end{center}

\vskip 1. cm
\begin{center}
{\bf\large M.~Ciuchini$^{(a)}$, G.~D'Agostini$^{(b)}$, 
E.~Franco$^{(b)}$, V.~Lubicz$^{(a)}$,
} \\
{\bf\large G. Martinelli$^{(b)}$, F. Parodi$^{(c)}$, P. Roudeau$^{(d)}$
 and A. Stocchi$^{(d)}$}
\end{center}
\vspace*{0.3cm}
\begin{center}
\noindent
{\small
\noindent
{\bf $^{(a)}$   Universit{\`a} di Roma Tre
 and INFN,  Sezione di Roma III,}\\  
\hspace*{0.5cm}{Via della Vasca Navale 84, I-00146 Roma, Italy}\\
\noindent
{\bf $^{(b)}$ Universit\`a ``La Sapienza'' and Sezione INFN di Roma,}\\
\hspace*{0.5cm}{Piazzale A. Moro 2, 00185 Roma, Italy}\\
\noindent
{\bf $^{(c)}$ Dipartimento di Fisica, Universit\`a di Genova and INFN}\\
\hspace*{0.5cm}{Via Dodecaneso 33, 16146 Genova, Italy}\\
\noindent
{\bf $^{(d)}$ Laboratoire de l'Acc\'el\'erateur Lin\'eaire}\\
\hspace*{0.5cm}{IN2P3-CNRS et Universit\'e de Paris-Sud, BP 34, 
F-91898 Orsay Cedex}}\\
%{\bf $^{(g)}$ CERN}\\
%\hspace*{0.5cm}{Geneva, Switzerland}\\
\noindent
\end{center}
\vskip .5cm
\begin{abstract}
\noindent
Within the Standard Model,
a review of the current determination of the sides and angles of the CKM
unitarity triangle is presented, using experimental 
constraints from
the measurements of $\epsilonk$, $\vubsvcb$, $\dmd$ and from the limit 
on $\dms$, available in September 2000.  Results from
the experimental search for ${B}^0_s-\bar{{B}}^0_s$ oscillations
are introduced in the present analysis using  the likelihood.
Special attention is devoted to the determination of 
the theoretical uncertainties. The purpose of the analysis is to 
infer  regions where the parameters of 
interest lie with given probabilities. 
The BaBar ``{\it 95\%\,C.L. scanning}" method is also commented. 
\end{abstract}
\vspace*{1cm}

\vspace{\fill}

\centerline { To be  submitted to JHEP }
\newpage
\pagestyle{plain}

\setcounter{page}{1}    

\section {Introduction}
\label{sec:1}
In the Standard Model,  weak interactions of quarks are governed 
by the four parameters of the CKM matrix~\cite{ref:ckm} which, 
in the Wolfenstein  parametrisation~\cite{ref:wolf}, 
are labelled as $\lambda,~A,~\rhobar~{\rm and}~\etabar$~\footnote{
$\rhobar=\rho (1-\frac{\lambda^2}{2})$ and $\etabar=\eta
 (1-\frac{\lambda^2}{2})$~\cite{ref:blo}.}.
Measurements of semileptonic decays of strange and beauty particles
are the main sources of information on $\lambda$ and $A$, respectively.
The values of $\epsilonk$, $\vubsvcb$, $\dmd$ and ${\dms}$
provide a set of four constraints for $\rhobar$ and $\etabar$. 
These constraints depend, in addition, on other quantities obtained 
from measurements and/or theoretical calculations. 
The regions of  $\rhobar$ and $\etabar$ preferred by the four constraints
are expected to overlap, as long as the Standard Model
gives an overall description of the various experimental observations. 

Since several years there has been intense activity to constrain the allowed
region in the ($\rhobar, \etabar)$ plane from 
the best knowledge of the experimental and theoretical 
inputs~\cite{ref:pion5}--\cite{ref:herab}. Though the
analysis methods differ in some details, they have a common 
ground in what we shall call  {\it standard approach} 
through this paper. First, the goal of the various authors 
has been, explicitly or implicitly, to infer regions in which 
the values of $\rhobar$ and $\etabar$ are contained with a certain
level of probability (or {\it confidence}). Second, 
uncertainties due to statistical errors  and systematic effects
in experiments, as well as 
theoretical uncertainties, are combined together to deduce a global 
uncertainty about $\rhobar$ and $\etabar$.  
As far as this second point is concerned, the various authors have used 
different ``prescriptions"  which can be seen, indeed, as approximations
of the consistent Bayesian method which is described, 
and adopted, in this paper.    
The 68\% probability regions favoured by the data and the theoretical  
understanding of the relevant processes select quite
narrow regions for $\rhobar$ and $\etabar$, largely independent of 
the details of the specific methods and of the different 
treatment  of (experimental) systematic and theoretical uncertainties.  
\par A different approach,  named ``{\it 95\%\,C.L. scanning}"
in this paper\footnote{Note that
this {\it 95\%\,C.L. scanning} is different from the
``scanning method'' used
to predict e.g. $\varepsilon^\prime/\varepsilon$~\cite{ref:scanning}
(see also~\cite{ref:sceptical} for comments).},
has been adopted in the BaBar Physics Book~\cite{ref:Bphysb},
and  recently used in~\cite{ref:plasz}.
In this approach, it is stated that it is not possible to define probability 
distributions for    theoretical parameters  coming from 
calculations  affected  by systematic uncertainties  or based on educated guesses 
(in practice all  theoretical parameters and some experimental systematics 
belong to this class).
On the basis of these considerations,
the {\it 95\%\,C.L. scanning} approach rejects the  two basic points of the
 standard method and a different procedure is proposed.
For the theoretical inputs,   it is assumed
that one can only define   intervals inside which the {\it true} 
values of the parameters are  contained.
At fixed values of the theoretical inputs (within the allowed
intervals) a maximum likelihood fit, which includes the other sources of
uncertainties, is made,  and  95\% C.L. contours  are determined. 
Finally, the envelope of  such contours  is ``{\it proposed to be
a (conservative) method to obtain some 95\% C.L. regions for
all CKM parameters''}~\cite{ref:plasz}. Using  the same  arguments of the BaBar Physics 
Book~\cite{ref:Bphysb}, this procedure has been  recently  recommended in~\cite{ref:falk,ref:stone},
as opposed  to the standard approach which is claimed of being too
optimistic.
\par  In view of the importance of constraining the parameters of the 
CKM matrix, or of the possibility of detecting signals of new physics 
in low-energy weak decays, in this paper we  reconsider the whole 
matter,  and in particular we focus  on the most 
critical issues of the CKM-triangle analysis, namely the uncertainty on  
theoretical parameters and the inferential framework to handle 
consistently all uncertainties.  
This also allows to answer to several important
controversial  questions raised 
in the past, namely:
\begin{itemize} 
\item  whether it is possible, or necessary, to assign a probability 
distribution function (p.d.f.) to theoretical parameters; 
\item  whether it is possible to   define a p.d.f. for quantities 
extracted from  physical measurements 
and from theoretical parameters affected by systematic uncertainties;
\item whether average values, errors and  p.d.f. of the quantities considered in the present analysis
depend in a crucial way on the assumptions made on the theoretical parameters and the corresponding errors. 
\end{itemize}
 
The main conclusions of this study are the following: 
\begin{enumerate} 
\item  The standard method is a theoretically
sound approach, which finds its justification within the inferential 
framework discussed in this paper. This method allows   a  consistent treatment of the systematic 
and theoretical uncertainties and makes it possible to define regions
where  the values of $\rhobar$ and  $\etabar$ (as well as 
of other quantities of phenomenological interest) are  contained with any given
level of confidence. In this 
respect  the criticisms of~\cite{ref:falk,ref:stone}
are  not justified.
\item   The BaBar {\it 95\%\,C.L. scanning}, instead, 
is based on an {\it ad hoc} prescription intended to define {\it only} 
a ``95\% C.L." region. The meaning of this 
statement  is unclear: 
the so called ``95\% C.L. region''  does not correspond
to the usual statistical definition  of 
95\% confidence   that the values of parameters lie
in that region, neither in a frequentist sense, nor in a Bayesian one.
\item  For the sake of comparison, with  the standard method
we have used the same values of  the input parameters  and 
 tried to mimic the same uncertainties of the {\it 95\%\,C.L. scanning}.
We find that the ``95\% C.L.'' regions selected  in the $(\rhobar,~\etabar)$
plane with the two methods
are very similar, and it is thus not founded to qualify as too optimistic
the standard approach.
\item In the {\it 95\%\,C.L. scanning} approach the information contained in the p.d.f.
of the relevant quantities ($\rhobar$, $\etabar$, $\sin (2 \beta)$, etc.)
 is missing.
Thus we only know that a certain quantity is somehow expected in 
a given interval, but
we do not know which is the most probable value, what is the shape of its
 p.d.f. etc. We show, instead,  that this important information can be extracted 
 from the
data, using the standard method, in spite of the uncertainties 
in the theoretical parameters.
\end{enumerate}

Regarding the analysis, since  this kind of  studies have been extensively
illustrated in previous
publications, see  for example~\cite{ref:parodietal} 
and~\cite{ref:roma}, here we only discuss in detail ${B}^0_s-\bar{{B}}^0_s$ mixing,
for which we adopted a different procedure, based directly on the likelihood
(Section~\ref{sec:deltams}).

The main  results for the physical quantities  ($\rhobar$,
$\etabar$, $\sin(2\beta)$, etc.)  can be found in Sections~\ref{sec:results}.
 
The remainder of the paper is organised as follows.
In Section~\ref{sec:formulae} we summarise the theoretical constraints
between  $\rhobar$ and $\etabar$ and the available experimental and 
theoretical inputs in the Standard Model. 
In Section~\ref{sec:inference} we describe the inferential 
framework used in this study and relate it to different
standard approach analyses.  Our comments  on  the {\it 95\%\,C.L. scanning}
method are also presented in this Section.
The choice of values and uncertainties
for the most critical theoretical parameters is discussed  
in Section~\ref{sec:parth}. In Section \ref{sec:other} the p.d.f. determination for $\Vcb$, $\Vub$ and
the parameter $\lambda$ are explained. A new  method to include the 
information coming from  searches of ${B}^0_s-\bar{{B}}^0_s$ mixing
is illustrated  in Section~\ref{sec:deltams}. 
The results of the analysis are presented and 
discussed in Section~\ref{sec:results}. 
The stability of the results has been verified by varying the different input parameters
in Section~\ref{sec:stability}. 
A comparison of our results with those obtained with {\it 95\%\,C.L. scanning}
is made in Section~\ref{sec:comparison}.  Finally, conclusions are drawn in Section~\ref{sec:conclusions}.

\section{Standard Model formulae relating \boldmath$\rhobar$ and \boldmath$\etabar$
to experimental and theoretical inputs}\label{sec:formulae}

Four measurements restrict, at present, the possible range of variations of 
the $\rhobar$ 
and $\etabar$ parameters:
\begin{itemize}
\item The relative rate of charmed and charmless $b$-hadron semileptonic 
decays 
which allows to measure the ratio
\begin{equation}
\vubovcb = \frac{\lambda}{1~-~\frac{\lambda^2}{2}}\sqrt{\rhobar^2+\etabar^2}\, .
\label{eq:C_vubovcb} \end{equation}.
\item The ${B}^0_d-\bar{{B}}^0_d$ time oscillation period
which can be related to the mass difference between the light and heavy 
mass eigenstates
of the ${B}^0_d-\bar{{B}}^0_d$ system
\begin{equation}
\dmd\ = 
\ {G_F^2\over 6 \pi^2} m_W^2 \ \eta_c S(x_t) \ A^2 \lambda^6 \
[(1-\rhobar)^2+\etabar^2] \ m_{B_d} \ f_{B_d}^2 \hat B_{B_d} \ , 
\label{eq:deltam} \end{equation}
where $S(x_t)$ is the Inami-Lim function~\cite{inami} and  $x_t=m_t^2/M^2_W$.
$m_t$ is the $\overline{MS}$ top mass,
$m_t^{\overline{MS}}(m_t^{\overline{MS}})$,  and 
$\eta_c$ is the perturbative QCD short-distance NLO correction.
The remaining factor, $f_{B_d}^2 \hat B_{B_d}$, encodes the information 
of non-perturbative QCD. 
Apart for $\rhobar$ and $\etabar$, the most uncertain parameter in this 
expression is $f_{B_d} \sqrt{\hat B_{B_d}}$. The value of $\eta_c=0.55 \pm 0.01$
has been obtained in~\cite{ref:bur1} and 
we used   $m_t=(167 \pm 5)\,  \GeV$, as
deduced from measurements of the  mass by CDF and D0 Collaborations
\cite{ref:top}.
\item
The limit on the lower value for the time oscillation period
of the ${B}^0_s-\bar{{B}}^0_s$ system is transformed
into a limit on $\dms$ and compared with $\dmd$
\begin{equation}
\frac{\dmd}{\dms}~=~\frac{m_{B_d}f_{B_d}^2 \hat B_{B_d}}
{m_{B_s}f_{B_s}^2 \hat B_{B_s} }\
\left( \frac{\lambda}{1-\frac{\lambda^2}{2}} \right )^2 \ [(1-\rhobar)^2+\etabar^2]\, .
\label{eq:dms} \end{equation}
The ratio
$\xi~=~f_{B_s}\sqrt{\hat B_{B_s}}/f_{B_d}\sqrt{ \hat B_{B_d}}$
is expected to be better determined from theory than the individual
quantities entering into its expression. In  our analysis, we accounted for the 
correlation due to the appearance of $\dmd$ in both Equations~(\ref{eq:deltam})
and (\ref{eq:dms}).
\item CP violation in the kaon system which is expressed by $\epsilonk$
\begin{equation}
\epsilonk\ = C_\varepsilon \ A^2 \lambda^6 \ \etabar
\left[ -\eta_1 S(x_c) 
+ \eta_2 S(x_t) \left( A^2 \lambda ^4 \left(1-\rhobar\right)\right)
+ \eta_3 S(x_c,x_t) 
 \right] \ \hat B_K  
\ , \label{eq:epskdef}
\end{equation}
where 
\begin{equation} C_\varepsilon = \frac{G_F^2 f_K^2 m_K m_W^2}
{6 \sqrt{2} 
\pi^2 \Delta m_K} \ . \end{equation}
$S(x_i)$ and $S(x_i,x_j)$ are the appropriate  
Inami-Lim functions~\cite{inami} of $x_q=m_q^2/m_W^2$, including the
next-to-leading order QCD corrections~\cite{ref:bur1,ref:basics}.
The most uncertain parameter is $\hat \BK$. 
\end{itemize}
Constraints are obtained by comparing present measurements with 
theoretical expectations using the expressions given above 
and taking into account the different sources of uncertainties. In addition
to $\rhobar$ and $\etabar$, these expressions depend on other quantities
which have been listed in Table \ref{tab:1}.
Additional measurements or theoretical
determinations have been used to provide information on the values of
 these parameters.

\begin{table}[htb!]
\begin{center}
\footnotesize{
\begin{tabular}{|c|c|c|c|c|}
\hline
                       Parameter                 &  Value &
 Gaussian &  Uniform & Ref. \\
 & &   $\sigma$ & half-width &  \\
\hline
$\lambda$    & $0.2237$ & \multicolumn{2}{|c|}{  $0.0033$\,, see text}   &sect. \ref{secvud} (eq. \ref{eq:lambda_corr}) \\
$\left | V_{cb} \right |$ & $41.0 \times 10^{-3}$
     & \multicolumn{2}{|c|}{ $1.6\times 10^{-3}$\,, see text} &sect. \ref{secvcb} (eq. \ref{eq:vcb_corr})\\

$\left | V_{ub} \right |$  & $ 35.5  \times 10^{-4}$ & $ 3.6 \times 10^{-4}$ & -- &sect. \ref{secvub} (eq. \ref{eq:vub_corr})\\
$\epsilonk$   &
$2.271 \times 10^{-3}$&
$0.017
 \times 10^{-3}$  & --
     &~\cite{ref:pdg00} \\
     $\Delta m_d$  & $0.487~\mbox{ps}^{-1}$ & $0.014~\mbox{ps}^{-1}$ & --
     &~\cite{ref:osciw}  \\
$\Delta m_s$  & $>$ 15.0 ps$^{-1}$ at 95\% C.L.
     & \multicolumn{2}{|c|}{see text} &~\cite{ref:osciw}  \\
$m_t$ & $167~\GeV$ & $ 5~\GeV$ & --
     &~\cite{ref:top} \\
$m_b$ & $4.23~\GeV$ & $ 0.07~\GeV$ & --
     &~\cite{ref:hispanicus} \\
$m_c$  & $1.3~\GeV $ & $0.1~\GeV$
     & --  &~\cite{ref:pdg00} \\
$\hat \BK$    & $0.87$ & $0.06$ &  $0.13$
     & sect. \ref{sec:parth} (eq. \ref{eq:lbk})     \\
$f_{B_d} \sqrt{\hat B_{B_d}}$ & $230~\MeV$  & $25~\MeV$
     &  $20~\MeV$  & sect. \ref{sec:parth} (eq. \ref{eq:fbn}) \\
$\xi=\frac{ f_{B_s}\sqrt{\hat B_{B_s}}}{ f_{B_d}\sqrt{\hat B_{B_d}}}$
     & $1.14$ & 0.04 & $ 0.05 $& sect. \ref{sec:parth} (eq. \ref{eq:xival}) \\
$\alpha_s$  &  $0.119$ & $0.03$ & -- &~\cite{ref:basics} \\
$\eta_1$  &  $1.38$ & $0.53$ & -- &~\cite{ref:basics} \\
$\eta_2$  &  $0.574$ & $ 0.004$ & --
     &~\cite{ref:bur1} \\
$\eta_3$  &  $0.47$ & $ 0.04$ & -- &~\cite{ref:basics} \\
$\eta_b$  &  $0.55$ & $ 0.01$ & --
     &~\cite{ref:bur1} \\
$f_K$     & $0.159~\GeV$ & \multicolumn{2}{|c|}{fixed}
     &~\cite{ref:pdg00} \\
$\Delta m_K$  & $
0.5301
\times 10^{-2} ~\mbox{ps}^{-1}$
     & \multicolumn{2}{|c|}{fixed} &~\cite{ref:pdg00} \\
 $G_F $   & $
 1.16639
\times 10^{-5} \GeV^{-2}$
     & \multicolumn{2}{|c|}{fixed}&~\cite{ref:pdg00} \\
$ m_{W}$  & $80.42
 ~\GeV $
     & \multicolumn{2}{|c|}{fixed} & ~\cite{ref:pdg00} \\
$ m_{B^0_d}$ & $5.2792
~\GeV $
     & \multicolumn{2}{|c|}{fixed} &~\cite{ref:pdg00} \\
$ m_{B^0_s}$ & $5.3693
~\GeV $
     & \multicolumn{2}{|c|}{fixed} &~\cite{ref:pdg00} \\
$ m_K$ & $0.493677
 ~\GeV$
     & \multicolumn{2}{|c|}{fixed} &~\cite{ref:pdg00} \\
\hline
\end{tabular} }
\label{tab:1}
\end{center}
\caption[]{ \small { Values of the quantities entering into the expressions
of $\epsilonk$, $\vubsvcb$, $\dmd$ and $\dms$.
In the third and fourth  columns the Gaussian and the flat part of the
uncertainty are given,  respectively. }}
\end{table}

\section{Inferential framework}\label{sec:inference}
In this Section we recall the basic ingredients of the
standard method, interpreted in the framework of the Bayesian approach.
This  allows us
to  discuss the role of the systematic and theoretical uncertainties
in deriving probability intervals for the relevant parameters.
\subsection{Standard approach and Bayesian inference}
Each of Equations~(\ref{eq:C_vubovcb})--(\ref{eq:epskdef})
relates a  constraint $c_j$
(where $c_j$ stands for $\vubsvcb$, $\dmd$,  
$\dmd/\dms$ and $\epsilonk$, for $j=1,\ldots, 4$) 
to the CKM--triangle parameters $\rhobar$ and $\etabar$, via 
the set of ancillary parameters ${\mathbf x}$, 
where ${\mathbf x} =\{x_1, x_2, \ldots, x_N\}$ stand for all 
experimentally determined or theoretically calculated 
quantities  from which the various $c_j$ depend
\begin{equation}
c_j=c_j(\rhobar,\etabar; {\mathbf x}). 
\label{eq:c_j}
\end{equation} 
In an ideal case of  exact knowledge of  $c_j$ and ${\mathbf x}$,
each of the  constraints  provides a curve in the 
$(\rhobar,\etabar)$ plane. In such a case, there would be no
reason to favour any of the points on  the curve, unless we have some 
further information or physical prejudice, which 
might exclude  points outside a determined {\it physical
region}, or, in general, assign different weights to  different points. 
In a realistic  case, we suffer from  several uncertainties  on the quantities  
$c_j$ and ${\mathbf x}$. Uncertainty does not imply, however,
that we are absolutely ignorant about a given  quantity. 
First of all, there are values which,
to the best of our knowledge, we consider ruled out  (for example 
a value of $m_t$ of 100 GeV or 500 GeV). Second, 
we assign different probabilities
to the  values within the ``almost certain range'',
$147\,  \mbox{GeV}  < m_t < 187\, \mbox{GeV}$ say~\footnote{
In this example $m_t$ is the
$\overline{MS}$ top mass of Equation~(\ref{eq:deltam}),
$m_t = (167 \pm 5)$ GeV.}.
In the $m_t$  case, for example, we think that it is  much more probable
that the value of ${m}_t$ lies  between 
157 and 177 GeV rather than in the  rest of the  interval, in spite of the fact
that   the two sub-intervals have the  same widths. 
\par
This means that, 
instead of a  single curve (\ref{eq:c_j})   in the $(\rhobar,\etabar)$ plane,
we have a family   of  curves  
which depends on the distribution of  the set $\{c_j,{\mathbf x}\}$.
As a result, the points  in
the $(\rhobar,\etabar)$ plane get  different weights
(even if they were taken to be  equally probable {\it a priori}) and  
our {\it confidence} on the values of 
$\rhobar$ and $\etabar$  clusters in a region of the plane. 

The above arguments, which we consider very natural
and close to physicist intuition, can be  formalised
by using the so called Bayesian approach (see~\cite{ref:YR9903}
for an introduction). In this approach, 
the uncertainty  is described in terms of 
a probability density function  $f(\cdot)$, which quantifies
our confidence on the    values of a  given   quantity. The inference 
of  $\rhobar$ and $\etabar$ becomes then a straightforward application
of probability theory, getting rid of all  
``{\it ad hoc} prescriptions". 

The simplest way to implement the probabilistic reasoning discussed above  is to
define  an  idealised ``p.d.f." for each constraint 
\begin{equation}
f(\rhobar, \etabar\,|\,c_j,{\mathbf x}) \propto
\delta(c_j-c_j(\rhobar,\etabar,{\mathbf x}))\,,
\label{eq:ideal}
\end{equation}
where $\delta$ is the Dirac delta distribution. 
The quotes recall us  that this  p.d.f. is  a distribution in a mathematical
sense,  which is to be taken  as  the limit of a very narrow 
 p.d.f. with  values different from zero only  along a curve. 
The p.d.f. which takes into account the full uncertainty
about $c_j$ and ${\mathbf x}$ 
is obtained from (\ref{eq:ideal}) by making use of the
standard probability rules 
\begin{eqnarray}
f(\rhobar, \etabar) &=& \int
f(\rhobar, \etabar\,|\,c_j,{\mathbf x}) \cdot 
f(c_j,{\mathbf x})\, \mbox{d}c_j \, \mbox{d}{\mathbf x} 
\label{eq:prop1}\\
&\propto & \int \delta(c_j-c_j(\rhobar,\etabar,{\mathbf x}))\cdot 
f(c_j)\cdot f({\mathbf x}) \, \mbox{d}c_j \, \mbox{d}{\mathbf x} 
\label{eq:prop3} \\
&\propto & \int  \delta(c_j-c_j(\rhobar,\etabar,{\mathbf x}))\cdot 
\frac{1}{\sqrt{2 \pi}\,\sigma(c_j)}
    \exp\left[-\frac{(c_j-
\hat{c}_j)^2}{2\,\sigma^2(c_j)}\right] \cdot 
 f({\mathbf x}) \, \mbox{d}c_j \, \mbox{d}{\mathbf x} \\
&\propto & \int \frac{1}{\sqrt{2 \pi}\,\sigma(c_j)}
    \exp\left[-\frac{(c_j(\rhobar, \etabar, {\mathbf x})-
\hat{c}_j)^2}{2\,\sigma^2(c_j)}\right]
     f(x_1)\cdot f(x_2) \cdots 
f(x_N)\,  \mbox{d}{\mathbf x}  \,, 
\label{eq:prop33}
\end{eqnarray}
where $\hat{c}_j$ is the experimental best  estimate of $c_j$, 
with  uncertainty $\sigma(c_j)$.
A  Gaussian distribution 
has been assumed  just for simplicity  and without lack of generality.
The joint p.d.f. 
$f(c_j,{\mathbf x})$ has been splitted  as a product of the 
individual p.d.f., assuming the independence of the different quantities, 
which is a very good approximation for the case under study. 

As alternative procedure, one may introduce  a 
global inference  relating  $\rhobar$, $\etabar$,
$c_j$ and ${\mathbf x}$. This is  followed by a second step where 
marginalization  is performed over those quantities which we are not 
interested to.  In this case, by making use of Bayes' theorem,  we obtain
\begin{eqnarray}
f(\rhobar,\etabar, c_j,  {\mathbf x }\,|\,\hat{c}_j) & \propto &  
f(\hat{c}_j\,|\,c_j,\rhobar,\etabar,{\mathbf x})\cdot 
f(c_j,\rhobar,\etabar,{\mathbf x}) \label{eq:bayes1}\\
& \propto & f(\hat{c}_j\,|\,c_j)\cdot f(c_j\,|\,\rhobar,\etabar,{\mathbf x})
      \cdot f({\mathbf x},\rhobar,\etabar) \label{eq:bayes3}\\
&\propto & f(\hat{c}_j\,|\,c_j)\cdot\delta(c_j-c_j(\rhobar,\etabar,{\mathbf x}))
 \cdot f({\mathbf x})\cdot f_\circ(\rhobar,\etabar)\,,
\label{eq:bayes33}\end{eqnarray} 
where $f_\circ(\rhobar,\etabar)$ denotes the {\it prior} distribution
as often used in the literature.
The various steps follow from  probability rules,
by assuming  the independence of the different quantities and 
 by noting that 
$\hat{c}_j$ depends on ($\rhobar,\etabar,{\mathbf x}$) only 
via $c_j$. This is true since  $c_j$ is  
univocally determined, within the Standard Model, from the values of 
 $\rhobar$, $\etabar$ and ${\mathbf x}$ (hence the limit to a 
delta function of its p.d.f.). 
We then recover   
Equation~(\ref{eq:prop33}): i) by assuming a Gaussian error
function for $\hat{c}_j$ around $c_j$; ii) 
by considering the various $x_i$ as independent;
iii) by taking a flat a priori distribution for $\rhobar$ and $\etabar$ and
iv) by integrating Equation~(\ref{eq:bayes33}) over $c_j$ and ${\mathbf x}$. 
Note that equiprobability 
of all points in the $(\rhobar, \etabar)$ plane was also implicit 
in Equation~(\ref{eq:prop3}), as  discussed above. 
 
Although the first derivation of  Equation~(\ref{eq:prop3}) is 
probably the most intuitive one, 
 hereafter we use the second one, which is  
the usual way of performing  Bayesian inference. The second procedure 
also shows  explicitly the connection with the methods which 
we denoted as  standard in the introduction.

The  extension of  the formalism to several 
constraints is straightforward. 
 We can rewrite Equation~(\ref{eq:bayes1}) as 
 \begin{eqnarray}
f(\rhobar,\etabar,{\mathbf x}\,|\,\hat c_1,...,\hat c_{\rm M})
&\propto &
\prod_{j=1,{\rm M}}f_j(\hat{c}_j\,|\,\rhobar,\etabar,{\mathbf x})
\times \prod_{i=1,{\rm N}}f_i(x_i) \times f_\circ(\rhobar,\etabar) \ .
 \label{eq:bayes_f}
\end{eqnarray} 
In the derivation of (\ref{eq:bayes_f}),
we have  used  the  independence of the different quantities.  
Moreover,  the   conditioning   from    
$f_j(\cdot)$ on the $c_j$  have been removed, since  the $c_j$ act as  intermediate
variables  which are finally integrated away. 
 The derivation (\ref{eq:prop1})--(\ref{eq:prop33})
can also  be easily extended to the case of several constraints and leads,
again, to the same result as  that found by using   Bayes' theorem. 
We have only to account properly 
the weight on  $c_j$,  induced  by the other constraint(s) 
previously considered. In this case Equation~(\ref{eq:ideal}) becomes 
$f(\rhobar, \etabar\,|\, c_1, \dots , c_M,{\mathbf x}) \propto
\delta(c_M-c_M(\rho,\eta,{\mathbf x}))\cdot 
f(\rho, \eta\,|\,c_1, \dots , c_{M-1}, {\mathbf x})$.

By integrating Equation~(\ref{eq:bayes_f}) over ${\mathbf x}$ 
we can rewrite the inferential scheme  in the
following convenient way
\begin{equation}
f(\rhobar,\etabar\,|\,{\mathbf \hat{c}},{\mathbf f})
\propto {\cal L}({\mathbf \hat{c}}\,|\,\rhobar,\etabar, {\mathbf f})
\times f_\circ(\rhobar,\etabar)\,,
\end{equation} 
where ${\mathbf \hat{c}}$ stands for the set of measured constraints,
 and
\begin{equation}
{\cal L}({\mathbf \hat{c}}\,|\,\rhobar,\etabar,{\mathbf f})
= \int 
\prod_{j=1,{\rm M}}f_j(\hat{c}_j\,|\,\rhobar,\etabar,{\mathbf x})
\prod_{i=1,{\rm N}}f_i(x_i)\, \mbox{d}{ x_i}
\label{eq:lik_int}
\end{equation}
is the effective overall likelihood which takes into account 
all possible values of $x_j$, properly weighted. We have
written explicitly that the overall likelihood depends 
on the best knowledge of all $x_i$, described by $f({\mathbf x})$. 

Whereas {\it a priori} all values for  $\rhobar$ 
and $\etabar$ are considered equally likely, 
{\it a posteriori}  the probability clusters 
around the point  which maximises the  likelihood. 
This is the reason why, in principle,  different
procedures for determining $\rhobar$ and $\etabar$,
based on the maximum likelihood,
are  equivalent to the method described here and should get 
similar  results.  We say ``in principle"  because  other methods are 
typically  implemented
using the  $\chi^2$ minimisation. This   implies the assumption of  
a multi-Gaussian solution of  the integral (\ref{eq:lik_int}), 
with overall  standard deviations  
which are simply a quadratic combinations
of the ``uncertainties" related to each $x_i$.
On the other hand, a quadratic combination relies 
on the approximative linear dependence of $c_j$ from the 
possible variations of $x_i$. 

In conclusion, the final (unnormalised) p.d.f. obtained 
starting from a flat distribution of $\rhobar$ and $\etabar$ is
\begin{equation}
f(\rhobar,\etabar) \propto 
 \int 
\prod_{j=1,{\rm M}}f_j(\hat{c}_j\,|\,\rhobar,\etabar,{\mathbf x})
\prod_{i=1,{\rm N}}f_i(x_i)\,\mbox{d}{x_i}\, .
\label{eq:flat_inf}
\end{equation}   
The integration can be done by Monte Carlo methods, the 
normalisation is trivial, and all moments can be calculated
in a (conceptually) easy way. Obviously there are several ways 
to implement the Monte Carlo integration, using different 
techniques to generate  events. 
A comparison of the results obtained with the approach  of~\cite{ref:roma},
where some effort  has  been done to improve 
the generation efficiency, and  the results of~\cite{ref:parodietal} 
is presented  in Section~\ref{sec:results}.

It is important to note that the inferential method does not  make any
distinction on whether the individual likelihood associated
to some  constraint is different from zero only in a narrow
region (and we usually refer to this case as ``measurement"), 
or it goes to zero only on one of the two sides 
(e.g. when  $c_j\rightarrow \infty$ or $0$). In the latter case,  
the data only provide an upper/lower bound to the value of the constraint. 
This is precisely what happens, at present, with $\Delta m_s$. Therefore, 
the experimental information about this constraint enters naturally
in the analysis (more details can be found in Section~\ref{sec:deltams}).
\subsection{Treatment of systematic and theoretical uncertainties}
At this  point, it is in order a discussion  on some   
important  ingredients of the analysis which raised some 
controversy in the past.  They are related to the quantitative
handling of the uncertainties due to systematic effects 
and to theoretical inputs. 

In Equation~(\ref{eq:prop3}) we have written 
explicitly that $\hat{c}_j$ is Gaussian distributed around $c_j$. 
As a consequence, we tend to say that  also $c_j$ is Gaussian distributed
around $\hat{c}_j$,  the inversion being  well understood 
in the case of random errors. The question is how to include the 
case where also systematic  uncertainties are present.  
One of the nice features of the Bayesian approach
is that the uncertainty has positively the same meaning, and there is no
conceptual distinction between the uncertainty  due to random fluctuations, 
which might have occurred in the measuring process, 
the uncertainty about the parameters of the theory,
and the uncertainty about  {\it influence quantities} 
(i.e. ``systematics") of not-exactly-known value, 
 (see~\cite{ref:YR9903} and 
\cite{ref:GdAMR}). Under the assumption that the individual likelihoods
$f_j(\cdot)$ of Equations~(\ref{eq:bayes_f})--(\ref{eq:flat_inf})
do only depend on random effects, the uncertainty due to 
systematics can be included using, again, 
concepts and formulae of conditional  probability. 
In fact, calling ${\mathbf h}$ the set of influence 
quantities on which the measured constraints
may depend, with joint p.d.f. $g({\mathbf h})$, the likelihood 
(\ref{eq:lik_int}) becomes
 \begin{equation}
{\cal L}({\mathbf \hat{c}}\,|\,\rhobar,\etabar,{\mathbf f},
{\mathbf g})
= \int 
\prod_{j=1,{\rm M}}f_j(\hat{c}_j\,|\,\rhobar,\etabar,{\mathbf x},
{\mathbf h})
\cdot f({\mathbf x}) \cdot 
g({\mathbf h})\, \mbox{d}{\mathbf x}\,\mbox{d}{\mathbf h}\,,
\label{eq:lik_int_syst}
\end{equation}
where we have written the p.d.f. of ${\mathbf x}$ in its general form,
allowing also correlations among the elements. 
As can be seen from Equation~(\ref{eq:lik_int_syst}), 
there is neither a conceptual nor a formal distinction between the
handling of ${\mathbf x}$  and ${\mathbf h}$. 
Therefore, we can simply extend the notation to include in ${\mathbf x}$
the influence parameters responsible of the systematic uncertainty, 
and use Equation~(\ref{eq:lik_int}) in  an extended way.
This is what it has been actually done in the past 
to infer   $f(\rhobar,\etabar)$ with 
the Monte Carlo integration method resulting from~(\ref{eq:flat_inf}). 
Moreover, see also~\cite{ref:GdAMR} and references therein, 
we arrive to the following 
conclusion. Irrespectively of the assumptions
made on the p.d.f. of  ${\mathbf x}$,
the overall likelihoods $f(\hat{c}_j)$ are approximately 
Gaussian because of  a mechanism similar to the central limit 
theorem (i.e. just a matter of combinatorics). 
This makes the results largely stable against variations within 
 {\it reasonable}
choices of models and parameters used to describe 
the uncertainties due to theory and systematics. 
This also  explains why methods 
based on $\chi^2$ minimisation (for example  refs.~\cite{ref:mele}, 
\cite {ref:AL} and \cite {ref:checchia})
can be considered as approximations
of the one used in refs.~\cite{ref:pprs},\cite{ref:parodietal},
\cite{ref:taipei} and~\cite{ref:roma}.
We stress again  that a common ground 
of all the methods that we classify
as standard is to produce  regions where  $\rhobar$ 
and $\etabar$ are contained with any given level of confidence.
On the contrary, the BaBar {\it 95\%\,C.L. scanning}
is based on an {\it ad hoc} prescription  which obscures
the meaning of the results.
In that approach, a so called ``95\% C.L.'' is produced,
which does not correspond to the usual 95\% confidence that
the parameters lie in that regions.

As far as the choice of the mathematical expression for $f_i(x_i)$
is concerned, practical examples of simple models can be found 
in ~\cite{ref:GdAMR}. Due to the  insensitivity 
of the result on the precise model, as  discussed
previously and as   shown later in this paper, 
we simplify the problem, by reducing the choice only to two possibilities. 
We choose a Gaussian model when the uncertainty is 
dominated by statistical effects, or  there are many
comparable  contributions to the systematics error, so that the 
central limit theorem applies. We choose a uniform
p.d.f. if the
parameter is believed to be (almost) certainly in a given interval, 
and the points inside this interval are considered 
 equally probable.
%
%In some cases, besides the analysis performed
%with a uniform distribution, a Gaussian model is also considered, with a $\sigma$
%equal to half of the total variation interval (this
%corresponds to a standard deviation increased by $\sqrt{3}$ as compared to the initial standard deviation of the
%flat distribution).
%This corresponds to assign to  the  interval of the uniform p.d.f. 
%a confidence  of about 92\%. 

A final comment,  before ending this Section, 
concerns the compatibility among  inferences provided
by  individual constraints. In the simplified approach
based on $\chi^2$ minimisation, a conventional evaluation 
of compatibility stems automatically from 
the  value of the $\chi^2$ at its minimum. 
This information is lost in the likelihood approach, but we accept 
this loss without any  regret, in view of what 
we gain. It is well known, indeed,  that crude
arguments about compatibility or incompatibility,  based only 
on the minimum value of the  $\chi^2$ and the number of degrees of freedom
lead to misleading conclusions (as premature claims of new 
physics have demonstrated in the past decades).
Therefore, we consider more reasonable to
judge  the compatibility of the constraints by comparing 
 partial inferences obtained when removing each constraint at the time.
 Examples are given in Figures~\ref{fig:bands} and \ref{fig:noepk}.
In our analysis, the overlap of the various constraints 
is excellent, 
and therefore we have no reason to suspect deviations from the 
Standard Model, given the available experimental information.
%%%%%%%%%%%%%%%%%%%%%%%%%%%%%%%%%%%%%%%%%%%%%%%%%%%%%%%%%%%%%%%%%
\section{Theoretical inputs:
\boldmath$\BK$, \boldmath$f_B \sqrt{\hat B_B}$ and the \boldmath$\xi$ parameter}
\label{sec:parth}
In this Section we discuss the theoretical inputs which have been used in
our phenomenological analysis.   In particular, we explain how 
central values and uncertainty  models
for the different quantities have been chosen.
This gives us the opportunity of clarifying some issues on which there is,
we think, some confusion in the literature.
This discussion may be instructive especially for non-lattice
experts (often experimentalists) who 
are engaged in this kind of analyses and have to find some orientation in
using results from   a {\it plethora} of lattice studies.

For example, in ref.~\cite{ref:herab}, the renormalization-scheme dependence of the meson
decay constants is mentioned. Unfortunately decay constants (which are related
to matrix elements of the weak axial current), as all measurable physical quantities,
are {\it scheme independent} by definition. Indeed the physical, scheme-independent axial current
is obtained from the lattice one by a finite renormalization constant,
$A_{\mu} = Z_{A} A_{\mu}^{latt}$, which can be (has been) determined
non-pertubatively with a negligible uncertainty~\cite{ref:bochicchio,ref:luscher}.
Scheme dependence only enters  some theoretical predictions
(but never in those for the decay constants) because the perturbative calculations
of the Wilson coefficients  in the effective Hamiltonians,
relevant  for weak decays and mixing, are truncated at a certain order (typically NLO).
This is  not a peculiarity 
of the lattice approach but it depends on  the limited number of orders which 
has been computed in continuum perturbation theory.
Moreover, the choice of presenting the calculations in $\overline{MS}$ is only a
traditional option which has not to do with the continuum or the 
lattice formulation of the theory. A discussion of this point in the case of $B^{0}$--$\bar B^{0}$ 
mixing can be found below.

Another source of confusion is the uncontrolled
propagation of  values and errors of the parameters
from one review talk to another, without verification on the original papers,
and  regardless of more recent calculations and progresses.
A typical example is the value $\xi \sim 1.3$. This number was only
found in~\cite{ref:bbs}. All other lattice calculations find 
for this quantity values between $1.11$ and 
$1.17$~\cite{ref:vittorio}--\cite{bernard2000}
in the quenched approximation and similar numbers  have been recently confirmed
by unquenched data~\cite{bernard2000}.
Similarly, in~\cite{ref:falk}, A.~Falk quotes  
$\xi = 1.14 (0.13)$ on the basis of a two-year old review by 
S.~Sharpe~\cite{sharpe98}.
In the absence of unquenched calculations,
the   rather generous uncertainty reported in~\cite{sharpe98}
was justified, at the time,   on the basis of   theoretical
estimates obtained by using quenched and unquenched chiral 
perturbation theory.  These estimates  have not been  confirmed 
by (partially) unquenched results,  which were already available last 
year~\footnote{ We  denote as {\it partially}
 unquenched results
those obtained with two sea-quark  flavours at values of the light-quark masses larger than the
physical ones, typically of the order of the strange quark mass,
or slightly below.}. Irrespectively of recent lattice progresses,
this very  large uncertainty, which is not supported by any explicit numerical  result,
risks to  survive and be used  in future phenomenological analyses.
 The most recent figures for the $B$-meson  decay constants, the $B$-parameters,
$f_{B_d} \sqrt{\hat B_{B_d}}$ and $\xi$ can be found in
Tables~\ref{tab:uno} and \ref{tab:due}.  These tables include
 the results presented at Lattice 2000~\cite{bernard2000}.
In the following, for all the theoretical parameters, we have used  results 
taken from lattice QCD.
There are several reasons for this choice, which has been adopted also in
previous studies of the unitarity 
triangle~\cite{ref:pion1,ref:pion4,ref:pprs,ref:parodietal,ref:roma}. 
Lattice QCD is not a model,
as the  quark model for example, and therefore physical quantities can be
computed from first   principles without arbitrary assumptions. 
It provides a method   for predicting all physical quantities
(decay constants, weak amplitudes, form factors) within  a unique, coherent
theoretical framework. For many quantities the statistical 
errors have been reduced to the percent level (or even less).
Although most of the  results are affected by systematic effects,
the latter can be ``systematically" studied and eventually corrected.
All the recent literature on lattice calculations is indeed focused on
discussions of the systematic errors and studies intended to reduce these
sources of uncertainty.  We are not aware of any other approach (1/N expansion,
QCD Sum Rules, etc.) where such a deep investigation of systematic
errors is being  carried out for a so large set of physical quantities 
as in lattice QCD.
Finally, in cases where predictions (non post-dictions) from lattice QCD
have been compared with experiments, for example 
$f_{D_s}$, the agreement has been found  very good.    

Obviously, for some quantities the uncertainty from lattice simulations
is far from being satisfactory and  further effort is needed
to improve the situation.  For the quantities considered here this is 
particularly true for  the $\Bd$--$\Bdb$ mixing amplitude, as discussed
below.
Nevertheless, for the reasons  mentioned before,
 we think  that lattice results and uncertainties  are  the most
reliable ones and  we have used them in our study.
\subsection{Statistical and systematic effects in lattice calculation uncertainties}
Lattice simulations are theoretical experiments carried out by numerical
integration of the functional integral by Monte Carlo techniques.
In this respect uncertainties are evaluated following criteria very close to
those used in experimental measurements.
The results are obtained with  ``statistical errors", i.e.
uncertainties originated by stochastic fluctuations,  which may be reduced
by increasing the sample of gluon-field configurations on which the  
averages are performed.  It is very reasonable
to assume that the statistical fluctuations have a Gaussian
(almost Gaussian) distribution. Hence, the probabilistic inversion needed to
infer the quantity of interest gives rise to Gaussian uncertainty models.

To convert the results of lattice simulations in predictions for the physical 
amplitudes several  steps are necessary: 
\begin{itemize}
\item[a)]  renormalization of the  relevant operators;  \par
\item[b)]   extrapolation to the continuum
limit, namely to zero lattice spacing ($a \to 0$);  
\item[c)]  unquenched calculations. The most precise numbers have been
obtained in the quenched approximation. Theoretical estimates and
some preliminary results in the (partially) unquenched case  are also known and they are
used to estimate the systematic errors of the quenched results.
\end{itemize}
a)--c) are the main sources of systematic errors  
for $\BK$, $f_{B_q} \sqrt{\hat B_{B_q}}$ and $\xi$ and are discussed 
separately in the following. 
Here we want only to stress that systematic errors from lattice calculations
are conceptually similar to some of the systematic errors 
present in  experimental measurements. 
\par Let us consider as an example 
discretization errors.  Since it is obviously impossible to work at zero 
lattice spacing,  a  method to correct for discretization 
effects is to compute a given physical quantity at different values of 
the lattice spacing and to extrapolate it to zero lattice spacing. The 
theory tells us whether the extrapolation has to be  linear or quadratic in 
$a$.  Thus for example  one fits a given quantity $Q$ as
\bea Q(a) = Q(a=0) + Q^{\prime} a + Q^{\prime\prime} a^{2} + \ldots 
\, ,  \eea
and takes $Q(a=0)$ as best estimate of the value of $Q$ in the continuum limit. 
Since the ``measurements" performed  at fixed lattice 
spacing are subject to a statistical error, the uncertainty in the 
extrapolated quantity is  inflated with respect to the points directly 
measured.  In the quenched case, extrapolations have been made for both $\hat B_{K}$, the 
$B$-meson decay constants and $\xi$. Systematic studies for the $B$-meson mixing 
parameters are still missing. From a comparison among
calculations performed by different groups at different lattice 
spacing  and with different lattice actions, an estimate of discretization errors
can be obtained, however,   also in these cases. 
\par  When an extrapolation to the continuum limit of the lattice data has 
been possible,  the final uncertainty results 
from the statistical error of the points measured at fixed lattice 
spacing (a residual uncertainty is present when linear or linear plus quadratic  
extrapolations  give different results). Thus, in this case, it is natural to assume
that  the final error has a Gaussian distribution.
\par As for the  errors coming from quenched calculations, partially 
unquenched calculations exist for several quantities considered in 
this study, namely $\hat B_{K}$ and  $f_{B_{d,s}}$. These 
calculations are usually performed with two light quarks in the 
fermion loops, at values of the light-quark  masses larger than the 
physical values and an extrapolation in these masses 
is required. The calculations are generally made at a fixed value of 
the lattice spacing and thus contain discretization errors.
An estimate of the quenching errors is obtained by comparing quenched
and unquenched  results  at similar  values of the lattice spacing. 
These comparisons are complemented by theoretical estimates of
this uncertainty obtained by using quenched and unquenched chiral 
perturbation theory techniques~\cite{sharpe96}.
\par The question arising at this point is: what is the best model
to  describe  the theoretical systematic errors in lattice calculations, 
and hence  the assessment of the uncertainty? 
We refer to the general introduction of the inferential 
framework given in Section~\ref{sec:inference}. 
As  happens with  systematic errors in experiments,
this   {\it completely} relies  on the confidence of 
the experts about  possible variations of an influence parameter, 
the effect of quenching or  what would happen in passing from a 
perturbative order to the other or changing the  renormalization scheme. 
These evaluations are unavoidably subjective,
though {\it not arbitrary}, as long as we use the 
judgements of  responsible experts for each input quantity. 
Using their judgements we commit ourselves too. Therefore, hereafter,
when we state that  a parameter lies in a certain range
with uniform distribution, it means that, in practice,  we are 
100\% confident that the parameter lies in that region, and that,
for any choice of a  sub-interval of half the width, we are in condition
of indifference (i.e. 50\% confidence) that the value of the 
parameter is inside the sub-interval or somewhere else. For a more 
extended discussion see for example~\cite{guru} and references therein.
\par   {\it In conclusion, we cannot  find
any conceptual difference which would force
us to treat experimental and theoretical uncertainties on a different footing
and
claim that the standard method is  a  perfectly justified scientific 
approach able to establish confidence levels for the quantities of interest. 
In the following, for each parameter, taken from lattice QCD evaluations, 
best estimates for its
central value and attached uncertainties are given. }

%To evaluate the effect of the choice of
%a Gaussian or a flat probability distributions on the uncertainties of fitted quantities, for
%$\hat \BK$ and $\fbdsqbd$ the flat part of the p.d.f. for systematic 
%uncertainties has been replaced by a 
%Gaussian distribution having a $\sigma$ equal to half of the total 
%variation interval. 
In order to check the stability of the  results, we have also
made the analysis with the flat part of the theoretical 
uncertainty increased by a factor two. 
 
\subsection{Evaluation of the parameter \boldmath$\fbdsqbd$}
Traditionally, the $B$-parameter of 
the renormalized operator  is defined as  \bea
\label{defB}
\langle  \bar {B}_{d} \vert   Q_d^{\Delta B=2} (\mu) \vert {B}_{d} 
\rangle 
= \frac{8}{3}\ m_{B_d}^2 f_{B_d}^2 \ B_{B_d}(\mu)\, , 
\eea
where $Q_d^{\Delta B=2}= (\bar b \gamma^\mu (1-\gamma_5) d)
(\bar b \gamma_\mu (1-\gamma_5) d)$ and 
$\mu$ is the renormalization scale.
This definition stems from the vacuum saturation approximation (VSA)
in which $B_{B_d}=1$. Similarly one can define $B_{B_s}$.
The renormalization group invariant
 $B$-parameter $\hat B_{B_d}$ of Equation~(\ref{eq:deltam}) is defined as
\bea
\label{bhat}
\hat B_{B_d} =  \alpha_s(\mu)^{-\gamma_0/2\beta_0}
 \left( 1 + {\alpha_s(\mu) \over 4 \pi} J\right) \  B_{B_d}(\mu) \ .
\eea
$\gamma_0=4$ in all  schemes whereas $J$ depends on the scheme used 
for renormalizing  $Q_d^{\Delta B=2} (\mu)$.
In the theoretical expressions, the physical amplitudes are always defined 
in terms of $\hat B_{B_d}$. The advantage is that this quantity is not only
renormalization scale, but also renormalization-scheme independent.
A residual  scheme dependence remains  only because the coefficient renormalizing the
lattice operator is computed at a fixed order in perturbation theory (NLO in this
case), whereas its matrix element is computed non-perturbatively. This problem
would arise in any approach that computes physical amplitudes by 
combining  the Wilson coefficients of
the effective Hamiltonian, which are computed perturbatively, with 
hadronic matrix elements. Thus  it is not specific to lattice calculations.
The scheme dependence can be reduced by increasing the order 
at which Wilson coefficients are computed in {\it continuum} perturbation 
theory.

Indeed the important quantity is not
the $B$-parameter itself but the combination $(\fbdsqbd)^2$
 which is used as an alias
for the physical amplitude to which it is simply related by the factor
$8 m_{B_d}^2/3$. This is similar to the kaon $B$ parameter. In that case,
however, the decay constant is taken from experiments. For the ${B}_d$ meson,
instead, we have to rely on theory also for the decay constant. Since
the calculation of $f_{B_d}$ and $\hat B_{B_d}$ are strongly correlated,
the best way is to take the combination $\fbdsqbd$ from a single 
calculation rather than using $f_{B_d}$ and $\hat B_{B_d}$ from different 
studies as often done in the literature.

The  most recent calculations of the combination $\fbdsqbd$,
in the quenching approximation, come from ref.~\cite{ape00}, 
obtained with the non-perturbatively improved action  and a non-perturbative
renormalization of the lattice  operators, and 
from ref.~\cite{ll2000}, with the mean-field improved action and
perturbatively renormalized operators
\bea \fbdsqbd  &=& (206 \pm28 \pm 7)\,{\rm MeV} \quad \cite{ape00}  \, ,
\nonumber \\ \fbdsqbd  &=& (211 \pm 21 ^{+27}_{-28})\,{\rm MeV}
\quad \cite{ll2000} \,  .\eea
These results correspond to the following values obtained in the same
simulation
\bea f_{B_d}& =& (174 \pm 22^{+7+4}_{-0-0}) \,{\rm MeV} \,  \quad
\hat  B_{B_d}   = 1.38 \pm 0.11 ^{+0.00}_{-0.09} \quad \cite{ape00} \, ,
\nonumber \\
f_{B_d}& =& (177 \pm 17^{+22}_{-26}) \,{\rm MeV} \,  \quad
\hat  B_{B_d}  = 1.41 \pm 0.06 ^{+0.05}_{-0.01}  \quad \cite{ll2000} \,
. \eea

\renewcommand{\arraystretch}{1.3}
\begin{table}
\begin{center}
\scriptsize{
\begin{tabular}{|c|c|c|c|c|c|}
\hline  \multicolumn{2}{|c|}{Quenched}& $f_{B_d}$ (MeV)&
$f_{B_s}/f_{B_d}$ &
$f_D$ (MeV)& $f_{D_s}$ (MeV) \\
\hline
APE~\cite{ape97} & 97& $180(32)$  & $1.14(8)$ & $221(17)$ &  $237(16)$
\\
FNAL~\cite{fnal97} & 97& $164(^{+14}_{-11})(8)$  & $1.13(^{+5}_{-4})$ &
$194(^{+14}_{-10})(10)$ & $213(^{+14}_{-11})(11)$ \\
JLQCD~\cite{jlqcd98} & 98& $173(4)(12)$  & $\simeq 1.15$ & $197(2)(17)$ &
$224(2)(19)$ \\
MILC$^{(*)}$~\cite{milc98} & 98& $157(11)(^{+25}_{-9})(^{+23}_{-0})$
& $1.11(2)(^{+4}_{-3})(3) $ & $192(11)(^{+16}_{-8})(^{+15}_{-0})$ &
$210(9)(^{+25}_{-9})(^{+17}_{-1})$ \\
APE~\cite{ape99} & 99& $173(13)(^{+34}_{-2})$ & $1.14(2)(1)$ & $216(11)
(^{+5}_{-4})$ & $239(10)(^{+15}_{-0})$ \\
APE~\cite{ape00} & 00& $174(22)(^{+7}_{-0})(^{+4}_{-0})$ &
$1.17(4)(^{+0}_{-1})$  & $207(11)(^{+3}_{-0})(^{+3}_{-0})$ &
$234(9)(^{+3}_{-0})(^{+2}_{-0})$ \\
UKQCD$^{(**)}$~\cite{ukqcd00} & 00 & $218(5)(^{+5}_{-41})$ & $1.11(1)(^{+5}_{-3})$ &
$220(3)(^{+2}_{-24})$ & $241(2)(^{+7}_{-30})$ \\
MILC~\cite{milc2000} & 00 & $173(6)(16)$ & $1.16(1)(2)$ &
$200(6)(^{+12}_{-11})$ & $223(5)(^{+19}_{-17})$ \\
CP-PACS~\cite{cppacs00} & 00 & $188(3)(9)$ & $1.148(8)(46)(^{+39}_{-0})$ &
$218(2)(15)$ & $250(1)(18)(^{+6}_{-0})$ \\
Lellouch and Lin~\cite{ll2000} & 00 & $177(17)(^{+22}_{-26})$ &
$1.15(2)(^{+3}_{-2})$ & $210(10)(^{+20}_{-16})$ & $236(8)(^{+20}_{-14})$ \\
\hline
\multicolumn{2}{|c|}{Unquenched}& $f_{B_d}$ (MeV)& $f_{B_s}/f_{B_d}$ &
$f_D$ (MeV)
& $f_{D_s}$ (MeV) \\
\hline
MILC~\cite{milc2000} & 00 & $191(6)(^{+24}_{-18})(^{+11}_{-0})$ &
$1.16(1)(2)(2)$ & $215(5)(^{+17}_{-15})(^{+8}_{-0})$ &
$241(4)(^{+32}_{-31})(^{+9}_{-0})$ \\
CP-PACS~\cite{cppacs00} & 00 & $208(10)(11)$ & $1.203(29)(43)(^{+38}_{-0})$ &
$225(14)(14)$ & $267(13)(17)(^{+10}_{-0})$ \\  \hline
\end{tabular} }
\caption[]{ \small {Decay constants from recent lattice calculations.
The errors are those of the original publications.  Some of the numbers have been taken from
the recent lattice reviews~\cite{ref:vittorio,ref:hashimoto}. ${(*)}$: By taking  
in the extrapolation  the previous MILC data closer to the
continuum limit only (corresponding to $\beta=6/g_0^2 \ge 6.0$), one would find $f_{B_d} \sim
180$ MeV.  This is, in our opinion, a better extrapolation of these data.
${(**)}$: This large value was found by using the Sommer parameter
to calibrate the lattice spacing.  Using the $\rho$ mass, instead,  they find $
f_{B_d}= 186$ MeV. We do not understand the origin of this large difference.}}
\label{tab:uno} 
\end{center}
\end{table} 

%%%%%%%%%%%%%%%%%%%%%%%%%%%%%%%%%%%%%%%%%%%%%%%%%%%%%%%%%%%%%%%%%%%%%%%%%%%
\renewcommand{\arraystretch}{1.3}
\begin{table}
\begin{center}
\footnotesize {
\begin{tabular}{|c|c|c|c|c|}
\hline \multicolumn{2}{|c|}{LQCD Calculations} & $B_{B_d}(m_b)$ & $\hat
B_{B_d}$
& $\hat B_{B_s}/\hat B_{B_d}$
\\ \hline
JLQCD~\cite{jlqcdxxx} & 96 & $0.85(6)$ & $1.26(8)$ & $\simeq 1$  \\
BBS~\cite{ref:bbs} & 98 & $0.96(12)$  & $1.42(18)$ & $\simeq 1$  \\
APE~\cite{ape00} & 00 &  $0.93(8)(^{+0}_{-6})$ & $1.38(11)(^{+0}_{-9})$
& $ 0.98(5)$  \\
Lellouch and Lin~\cite{ll2000} & 00 &  $0.95(4)(^{+3}_{-1})$ &
$1.41(6)(^{+5}_{-1})$ & $ 0.99(2)(^{+0}_{-1})$  \\
\hline
\multicolumn{2}{|c|}{HQET}& $B_{B_d}(m_b)$ & $\hat B_{B_d}$ &
 $\hat B_{B_s}/\hat B_{B_d}$  \\ \hline
Gimenez and Reyes~\cite{gr}(APE data) & 98 & $0.87(5)(3)(2)$ & $1.29(8)(5)(3)$ &
$--$ \\
Gimenez and Reyes~\cite{gr}(UKQCD data) & 98 & $0.85(4)(3)(2)$ & $1.26(6)(5)(3)$ &
$--$ \\  \hline
\end{tabular} }
\caption[]{ \small {$B$-meson $B$ parameters  from
recent lattice calculations.
For comparison we have evolved results and errors of the
original publications to a common scale.}}
\label{tab:due} 
\end{center}
\end{table} 
 
The numbers above agree with (our) world averages of quenched determinations, 
based on the results given in Tables~\ref{tab:uno} and \ref{tab:due},
which were
used in~\cite{conferences}~\footnote{ The same quenched results
for $f_{B_d}$ is quoted  in ref.~\cite{bernard2000}.}
\bea f_{B_d}& =& (175 \pm 20 )\,{\rm MeV} \,  \quad
\nonumber \\
\hat  B_{B_d} &=& 1.36 \pm 0.08 \pm 0.05 \, , \eea
which can be combined to give
\bea \fbdsqbd =  (205 \pm 24 \pm 8) \,{\rm MeV} \,  . \label{fbun}\eea
Note that in this case we have used decay constants and $B$-parameters
from different calculations.

\par Our average of $f_{B_d}$ does not include the results
obtained with NRQCD. The reason is that the quenched
results  from two different groups are incompatible between each other:
ref.~\cite{collins} found $f_{B_d}=147 (11)(^{+8}_{-12})(9)(6)$ MeV
to be contrasted with the recent CP-PACS result~\cite{bernard2000}
$f_{B_d}=191(5)(11)$ MeV. Moreover  \cite{collins} finds a
rather large difference,
of about $40$ MeV (from $147$ MeV $\to 186$ MeV),  between the
quenched and unquenched case. This  is not confirmed by the most recent results
given in Table~\ref{tab:uno}, which give differences of the order of $20$ MeV.
In our opinion,  the situation with  this approach is
still rather confused  and therefore we do not use the NRQCD
results (until it will not be clarified). This applies also
to the related calculations of the $\hat B_{B_{d,s}}$
parameters, which are computed within the same framework.

\par Our  average for $\hat B_{B_d}$ has been  computed by combining
values  obtained  with  heavy
quark masses in the charm region, extrapolated to the $B$ mesons 
(denoted as LQCD in Table~\ref{tab:due}), with those obtained  
in  the HQET~\cite{gr}, at  lowest order in the $1/m_b$ expansion. 
The systematic error has been estimated from the
different values obtained by using only  the LQCD results or by combining
 LQCD and HQET predictions.

The world average given above for $f_{B_d}$ includes data 
obtained after extrapolation to the continuum,
as well as results obtained with improved actions at small values
of $a$ (for which discretization errors are expected to be smaller).
A completely non-perturbative determination
of the axial current renormalization constant $Z_A$ 
has been also performed in several cases, thus eliminating 
this source of errors.
 
In the case of the $B$ parameter, no systematic continuum extrapolation has been 
attempted yet. In this case it is reassuring that results obtained 
with perturbative and non-perturbative renormalization techniques and for a variety
of values of lattice spacing are so close.  In the absence of any indication
of large discretization errors, we  ignore them in the following.

As far as quenching errors are concerned, there is a general agreement that the value of the 
decay constants increases in the unquenched case. This is supported by  both theoretical
estimates  with quenched and unquenched chiral perturbation 
theory~\cite{sharpe96},  and by explicit numerical calculations, see Table~\ref{tab:uno}.  
The MILC~\cite{bernard2000} and CP-PACS~\cite{cppacs00}
Collaborations find very consistent
results, 
\begin{eqnarray}
f_{B_d}^{unq}/f_{B_d}^{quen} & =& 1.12^{+0.16}_{-0.11}  \, , \nonumber \\
f_{B_d}^{unq}/f_{B_d}^{quen} & = & 1.11 \pm 0.06\, , \label{eq:rafb}  \end{eqnarray}
respectively.
Most unfortunately no unquenched determinations of the $B$-parameters, or even
better of $f_B \sqrt{\hat B_B}$, have been presented yet.
For the $B$-meson $B$ parameters, chiral perturbation theory suggests that the
unquenching error is at most of the order of 10\%.
 This prediction should be supported by explicit numerical simulations which are
missing at the moment.
\par  By assuming  10\% quenching uncertainty
for $\hat B_{B_d}$, and using the results 
in (\ref{fbun})  and (\ref{eq:rafb}), we arrive to
\begin{equation} \fbdsqbd= (230 \pm 25 \pm 20) \, \mbox{MeV}  . \label{eq:fbn} \end{equation}
which has been used in our analysis.  In our preliminary analysis
(mostly based on quenched results) which has
been presented in~\cite{conferences},  we used instead 
\begin{equation} \fbdsqbd=(220\pm 25 \pm 20 )\MeV \, ,\label{fbo}\end{equation}
which is very close to the present value. The changes 
in the results for 
the relevant quantities  ($\sin(2\alpha)$, $\sin(2\beta)$, $\gamma$, etc.),
induced by the difference between (\ref{eq:fbn}) and (\ref{fbo}),
are negligible.

\subsection{Evaluation of the parameter \boldmath$\xi$}
\par Besides the ${B}^0-\bar{{B}}^0$ amplitude,
an additional constraint is given by the 
ratio \bea \label{ratiom} \frac{\Delta m_s}{\Delta m_d} =\frac{\vert V_{ts}\vert^2}
{\vert V_{td}\vert^2} \frac{m_{B_s}}{m_{B_d}} \xi^2,\eea
%= \frac{\vert V_{ts}\vert^2} {\vert V_{td}\vert^2} r_{sd} \ ,\eea
where $\xi= f_{B_s}\sqrt{ \hat B_{B_s} }/f_{B_d}\sqrt{ \hat B_{B_d} }$.
By combining the results for the decay constant ratios $f_{B_s}/f_{B_d}$ 
and for the $B$ parameters (the latter being always very close to one)
(see Tables \ref{tab:uno} and \ref{tab:due} respectively)
one finds always a number of $\sim 1.15$.  As mentioned before, there is no
confirmation, neither in the quenched nor in the unquenched case, of a value
as large as $\xi=1.3$ as found in~\cite{ref:bbs}. 
With an estimate of the 
uncertainty on  $\hat B_{B_{s}}/\hat B_{B_{d}}$ of 10\%~\cite{sharpe96},
  we then find  
\begin{equation} \xi=1.14 \pm 0.03 \pm 0.05 \label{eq:xival} \end{equation}  
which is the value  used in our study.  In our preliminary analysis we used,
instead,
$\xi=1.14 \pm 0.06$~\cite{conferences}.

\subsection{Evaluation of the parameter \boldmath$\BK$}

The kaon $B$-parameter, $\BK$,  is one of the most studied, and more accurately
known,  quantities in lattice calculations. 
Very precise values   have been obtained, within the quenching 
approximation
\begin{eqnarray}
&& \BK^{\overline{\rm MS}}(\mu=2 \, {\rm GeV}) = 0.63 \pm 0.04 \quad 
\cite{ref:aokir} \nonumber \\
&& \BK^{\overline{\rm MS}}(\mu=2 \, {\rm GeV}) = 0.62 \pm 0.03  \quad 
\cite{gks}
\end{eqnarray}
which correspond to the renormalization group invariant $B$ parameter
\begin{equation}
\hat \BK= 0.87 \pm 0.06 \, .
\label{eq:bkhat}
\end{equation}
This uncertainty, meant as standard deviation, includes the 
contribution from the statistical error and the deviation from the extrapolation 
to the continuum limit (this error is much larger than those on individual data
at fixed lattice spacing).
The physical amplitude has been computed at the NLO by using 
boosted perturbation theory \cite{lpm}. A non-perturbative renormalization of the
lattice operator is preferable but has not been performed yet.
Previous experience on other quantities leads to an estimate of the error 
due to the use of the perturbative renormalization of the order of 5\%.
  
Uncertainties, due to the quenching approximation, have been evaluated
both theoretically and numerically.
Using chiral perturbation theory, the error
on $\BK$ due to the quenched approximation has been estimated 
to be negligible for degenerate quark masses ($m_s=m_d$) and of the 
order of 5\% for realistic quark masses~\cite{sharpe96}.
A numerical unquenched calculation with $n_f=2$ and $n_f=4$ resulted in a 
shift upwards of the value of $\BK$ by $(5\pm 2)\%$~\cite{kil}.
This calculation was performed at fixed lattice spacing and only the
difference between the quenched and unquenched $\BK$ for similar values of
$a$ was studied.  Both the theoretical estimate and the numerical evaluation
give a quenching error of ${\cal O}(5$\%$)$.  The residual uncertainty 
is due to our ignorance on the dependence of  this difference on the value of the lattice spacing. Since so far we have 
very accurate results with continuum extrapolation in the quenched case only
and unquenched results without continuum extrapolation,
it seems reasonable to consider a systematic uncertainty corresponding to a uniform distribution
spanning the range of 
$\pm 15$\% corresponding to the  maximum discretization effect
 expected in the unquenched case~\cite{ref:aokir}.
For the central value, we have taken the quenched result given in 
Equation~(\ref{eq:bkhat}).
We thus obtain
\begin{equation}
\hat \BK = 0.87 \pm 0.06 \pm 0.13
\label{eq:lbk} \end{equation}
which  has been used in our study.  This range of values is in good agreement with others
which can be found in the literature~\cite{ref:gupta,ref:lellouch}. 
The allowed  range for $\hat \BK$ in our evaluation and in 
\cite{ref:gupta,ref:lellouch} 
is smaller than  the one quoted  in~\cite{ref:Bphysb,ref:falk}, namely
$0.6 \le \hat \BK \le 1.0$.  That estimate tries to include results obtained
with techniques
different from lattice calculations, such as QCD Sum Rules and $1/N$ expansion.
In  our opinion, the other approaches lack of the accuracy 
and control of systematic effects reached by  lattice calculations for this 
parameter. We think instead that 
Equation~(\ref{eq:lbk})
corresponds to our best knowledge of $\hat \BK$, given the present
understanding of the theory.

\section{Other inputs}
\label{sec:other}
In this section we briefly discuss other  inputs which have been used 
in the present analysis: $\Vcb$, $\Vub$ and  $\lambda$. These are obtained
from  experimental measurements combined with several   theoretical 
predictions which are discussed below.
\subsection{Extraction of \boldmath$\Vcb$}
\label{secvcb}
Using exclusive decays $\Bdb \rightarrow \Dstarp \ell^- {\bar \nu_{\ell}}$,
the value of $\Vcb$ is obtained by measuring the differential decay rate at 
maximum $q^2$, which is the mass squared of the charged lepton-neutrino system.
At $q^2=q^2_{max}$, the $\Dstarp$ is produced at rest in the $\Bdb$ hadron
rest frame and HQET can be invoked to obtain the value of the corresponding
form factor $F_{\Dstar}(w=1)$. The variable $w$ is usually introduced as 
the product of the four-velocities of the $\Bdb$ and $\Dstarp$ mesons
\begin{equation}
w=v_{\Bdb}\cdot v_{\Dstarp}=
\frac{m^2_{\Bdb}+m^2_{\Dstar}-q^2}{2 \, m_{\Bdb} \, m_{\Dstar}}, \quad 
\quad w=1~{\rm for}~q^2=q^2_{max}\, . 
\label{eq:c_w}
\end{equation} 
In terms of $w$, the differential decay rate can be written as
\begin{equation} 
\frac{d BR_{D^{\ast}}}{dw}~=~\frac{1}{\tau_{B^0_d}} \frac{G_F^2}{48 \pi^3}m^3_{D^{\ast}}(m_B-m_{D^{\ast}})^2
K(w) \sqrt{w^2-1} F_{D^{\ast}}^2(w) \Vcb^2,
\label{eq:2.02}
\end{equation}
where $K(w)$ is a kinematic factor.

As the decay rate is zero for $w=1$, the $w$ dependence has to be adjusted over the measured range.

Four measurements obtained by the LEP collaborations \cite{ref:vcblep}
have been averaged with the result from CLEO \cite{ref:cleovcb}, taking into account correlations 
induced by common sources of systematic uncertainties.
%and using expressions more elaborate than (\ref{eq:2.02}).
Before averaging, results and errors  have been recalibrated using a common set of 
values for the external parameters, such  as the $\Bdb$ lifetime 
and charm hadron decay branching fractions \cite{ref:vcblep}.

The average value is 
\begin{equation}
F(1) \, \Vcb = (37.2 \pm 1.5)\times 10^{-3}, \quad \chi^2/NDF=9.15/4 \, .
\end{equation} 
The ``fit probability''\footnote{ ``p-value'' would be the correct modern statistics
term to be used instead of ``fit probability'' or ``$\chi^2$
probability"~\cite{ref:nn1,ref:nn2}. These expressions can be highly misleading as explained 
in \cite{ref:YR9903}.} is only 6\%, giving rise to the 
suspect that there could be some extra systematic effect playing 
an important role. 
\par Another evaluation of this average has thus been done following a 
model \cite{ref:dago1}
developed to combine results which appear to be in mutual disagreement.
It consists in assuming that quoted uncertainties ($s_i$) for 
each measurement are proportional to the unknown real uncertainty
($\sigma_i/r_i=s_i$). A distribution probability for $r_i$ is then assumed.
A simple model, which depends on two parameters, $\delta$ and $\zeta$ is given by
\begin{equation}
f(r)=\frac{2 \zeta^{\delta} r^{-(2\delta+1)}\exp^{-\zeta/r^2}}
{\Gamma(\delta)}\, .
\end{equation} 
%Provided that the mean value and standard deviation of $r$
%are taken of order 1, the final results are almost independent
%of  the  peculiar choice of values for the 
%two parameters. In the following $\zeta=0.6$ and $\delta=1.3$ have been used.
The natural choice for $\delta$ and $\zeta$ corresponds to an expected value of 1 for $r_i$, with 100\%
uncertainty. The final results are largely independent from the precise value of the
parameters. We have chosen the values of $\zeta=0.6$ and $\delta=1.3$ taken from Ref. \cite{ref:dago1}.

The probability distribution for the quantity $F(1) \, \Vcb$ is then obtained
using the Bayes theorem
\begin{equation}
f(x) \propto \int_{0}^{\infty}
\exp \left[
-\frac{1}{2}\sum_{i,j=1}^{5}(x-x_i)w_{i,j}
(\sigma_i/r_i,\sigma_j/r_j)(x-x_j) \right]  \left( \prod_{i=1}^{5} f(r_i)\, dr_i
\right)\, .
\label{eq:bayesvcb}
\end{equation} 
In this expression, $x=F(1)\, \Vcb$ and $w_{i,j}$ is the
weight matrix. The latter  is
 obtained by inverting the error matrix, after the 
inclusion of the  uncertainties given by the quantities $r_i$ 
(considered
to be independent for the different measurements). From this distribution,
which has non-Gaussian tails,
the average and the standard deviation have been obtained
\begin{equation}
F(1) \, \Vcb = (37.1 \pm 1.9)\times 10^{-3}.
\end{equation} 
The central value is practically the same as obtained with the usual fit and the standard deviation
has increased by 30\%.

From the exclusive  measurements and using  $F(1)=0.88 \pm 0.05$ 
\cite{ref:summer99}, $\Vcb$ has  been obtained
\begin{equation} \Vcb = (42.2 \pm 2.1 \pm 2.4) \times 10^{-3} \, . 
\end{equation} 

\begin{figure}[htb!]
\begin{center}
{\epsfig{figure=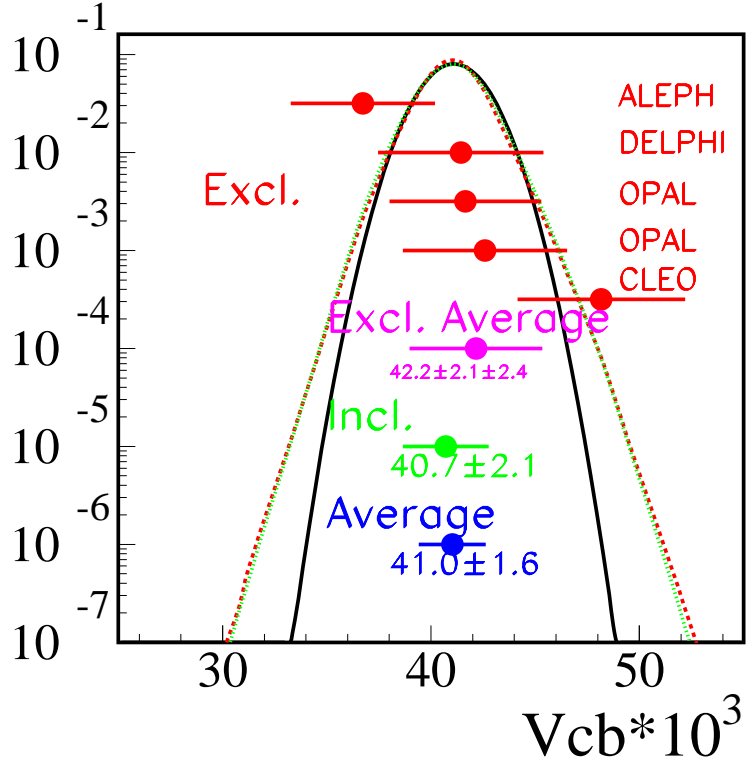,bbllx=30pt,bburx=540pt,bblly=30pt,bbury=540pt,height=10cm}}
\end{center}
\caption{ \small { Probability distribution function for $\Vcb$ obtained by
combining exclusive and inclusive measurements of $b$-hadron 
semileptonic decays. The full line distribution corresponds to a Gaussian
whereas dotted curves are obtained using the procedure explained in the text.
Two sets of values for the $\zeta$ and $\delta$ parameters, namely
(0.6, 1.3) and (1.4, 2.1) have been tried and correspond to almost 
identical distributions (dotted curves). The quoted error bars for individual results
correspond to all sources of uncertainties but
the combination procedure accounts for correlations, coming from common
systematic uncertainties.}}
\label{fig:vcball}
\end{figure}

The inclusive measurements of the semileptonic branching fraction
of $b$-hadrons give instead
\begin{equation}
\Vcb = ( 40.7 \pm 0.5 \pm 2.0 ) \times 10^{-3} \, .
\end{equation} 
The interested reader 
may consult \cite{ref:summer99} for more details on the 
quoted uncertainty for inclusive decays.

The combination of all results, using the procedure explained previously when averaging 
$F(1) \, \Vcb$ measurements, taking into account correlated systematics, gives
\begin{equation}
\Vcb = (41.0 \pm 1.6)\times 10^{-3} \, .
\label{eq:vcb_corr}
\end{equation}
The corresponding p.d.f, which has been used in the present analysis, 
is shown in Figure \ref{fig:vcball}.

\subsection{Extraction of \boldmath$\Vub$}
\label{secvub}

%%%%%%%%%%%%%%%%%%%%%%%%%%%%%%%%%%%%%%%%%%%%%%%%%%%%%%%%%%%
\begin{figure}
\begin{center}
{\epsfig{figure=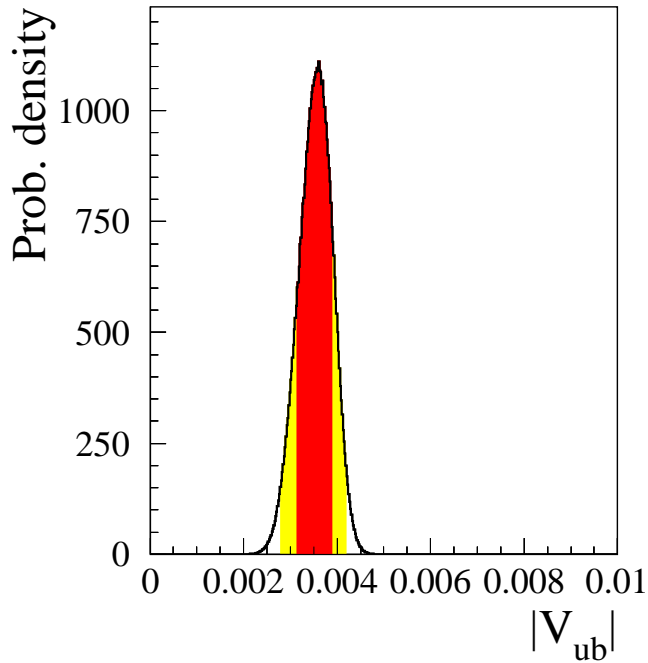,bbllx=1pt,bburx=240pt,bblly=1pt,bbury=240pt,height=9cm}}
\caption{ \small{The p.d.f. for  $\Vub$.}}
\label{fig:vubdist}
\end{center}
\end{figure}
%%%%%%%%%%%%%%%%%%%%%%%%%%%%%%%%%%%%%%%%%%%%%%%%%%%%%%%%%%%

$\Vub$ has been obtained from the measurements done at
CLEO and LEP.
 The CLEO collaboration~\cite{ref:vubcleo} has measured the branching fraction 
for the decay
$\bar{{B^0_d}} \rightarrow \rho^+ \ell^- \bar{\nu_{\ell}}$ and 
deduced a value for $\Vub$ using several models to describe the decay
form factors. LEP collaborations~\cite{ref:vublep} 
have developed dedicated algorithms to be sensitive
to a large fraction of the inclusive decay rate 
$b \rightarrow u \ell^- \bar{\nu_{\ell}}$ and, with some assumptions, 
%using models based on the Operator Product Expansion, 
a value for $\Vub$ is obtained.
The two measurements are 
\begin{eqnarray}
 \Vub &=& ( 32.5 \pm 2.9 \pm 5.5) \times 10^{-4}  \quad \mbox{CLEO}\nonumber \\
 \Vub &=& ( 41.3 \pm 6.3 \pm 3.1 ) \times 10^{-4}  \quad \mbox{LEP} \,
 \end{eqnarray}
where the second uncertainty is theoretical. The p.d.f. for the CLEO 
measurement is thus a convolution of a Gaussian and a flat distribution.
The theoretical error for the LEP measurement, being the convolution of several
different errors, is taken from a Gaussian distribution \cite{ref:vublep}.
Combining the two distributions,  in practice we obtain almost a Gaussian 
p.d.f. (see Figure \ref{fig:vubdist}) corresponding to  
\begin{equation}
\Vub = (35.5  \pm 3.6) \times 10^{-4}.
\label{eq:vub_corr}
\end{equation} 

\subsection{The parameter \boldmath$\lambda$}
\label{secvud}

\begin{figure}[htb!]
\begin{center}
{\epsfig{figure=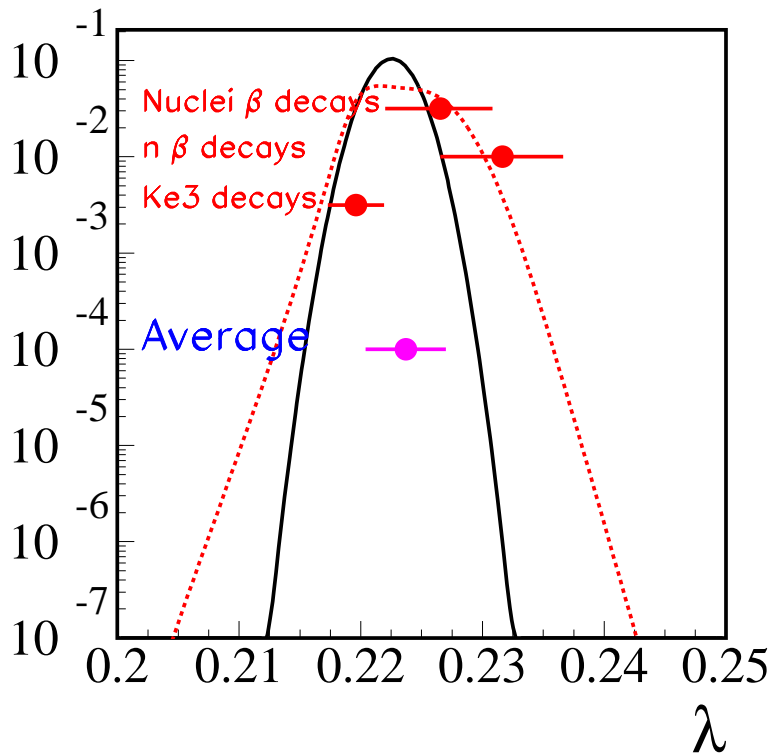,bbllx=30pt,bburx=540pt,bblly=30pt,bbury=540pt,height=10cm}}
\end{center}
\caption{\small { Probability distribution function for $\lambda$
obtained by combining measurements of $\Vud$ and $\Vus$.
The full line distribution corresponds to a Gaussian
whereas the dotted curve is obtained using the procedure explained in the text. }}
\label{fig:cabibbo}
\end{figure}

Measurements of $\Vud$ and $\Vus$ reported in \cite{ref:pdg00} have been 
combined \footnote{The corresponding values used in the present analysis are :
$\left|V_{ud}\right|$ = (0.9740 $\pm$ 0.0010) from nuclear beta decays, 
$\left|V_{ud}\right|$ = (0.9728 $\pm$ 0.0012) from neutron decay and
$\left|V_{us}\right|$ = (0.2196 $\pm$ 0.023) from $K_{e3}$ decays.} 
using the procedure explained in Section \ref{secvcb}, assuming
that $\Vud =\cos{\theta_c}$ and $\Vus=\sin{\theta_c} \equiv \lambda$. The additional
contribution from $\Vub^2$ to the unitarity condition can be safely neglected
owing to present uncertainties. The p.d.f. for $\lambda$ 
strongly deviates from a Gaussian (see Figure \ref{fig:cabibbo}) because present determinations of
this quantity from $\Vud$ and $\Vus$ measurements are more than two
standard deviations away. The average and the standard deviation of the obtained
distribution correspond to
\begin{equation}
\lambda = 0.2237 \pm 0.0033
\label{eq:lambda_corr}
\end{equation}
whereas a classical fit gives $\lambda = 0.2225 \pm 0.0019$.\\
Using the values of $\Vcb$ and $\lambda$ given respectively in Equations (\ref{eq:vcb_corr}) and
(\ref{eq:lambda_corr}) the following result is obtained for the parameter A
\begin{equation}
A = \frac{\Vcb}{\lambda^2} = 0.819 \pm 0.040
\label{eq:a_corr}
\end{equation}
It may be noticed that, in the present determination of the $\rhobar$ and $\etabar$ 
parameters, $\Vcb$ and $\lambda$, but not A, are used as constraints.

\section{A new procedure to include the information coming
from \boldmath${B}^0_s-\bar{{B}}^0_s$ oscillation analyses}
\label{sec:deltams}
In this Section, we illustrate a new procedure to include in the
analysis the experimental searches for ${B}^0_s-\bar{{B}}^0_s$ mixing.
It makes use of the likelihood along the lines discussed in
Section~\ref{sec:inference}.
%{\bf Roma-begin  in figs. 1c) and 2c) P(A) -> F(A) and F, A and R should be
% all 
%calligraphic (are you able to do this ??). 
%Roma-end }

The study of $\Bs-\Bsb$ oscillations is done by introducing the oscillation
amplitude ${\cal A}$ as  explained in~\cite{moser}.
Events have been distributed in two 
classes corresponding to oscillating and non-oscillating candidates,
according to the information obtained by tagging the presence
of a $B$ or a $\bar{B}$ meson at the beam interaction time and
at the $b$-hadron decay time. The expected decay-time distribution
of events is obtained by using a function which contains time 
distributions for all components, taking into account the non-perfect 
event classification, the different behaviour of $b$-hadron decays,
which depend on their type ($\Bdb$, $\Bm$, $\Bsb$ or $b$-baryon), and
background components originating from charm and light flavours.
The time distribution for the oscillating (non-oscillating) signal
is given by  
\begin{equation}
{\cal P}(\Bs \rightarrow \Bsb (\Bs) ) = \frac{1}{2\tau_{\Bs}}(1-
(+) {\cal A}\cos{\Delta m_s t})
e^{-t/\tau_{\Bs}} \, . 
\end{equation}
Such theoretical time distributions have been convoluted with the 
expected time 
resolution of the measurements. For each value of $\dms$ and each
analysis, 
the oscillation amplitude ${\cal A}$ is fitted. Fitted amplitude values 
have been combined \cite{ref:osciw} using a common set of 
determinations for external parameters and taking into account 
possible correlations between the different analyses. In this respect
the use of the oscillation amplitude provides a simple 
framework to account for all these effects.
The 95$\%$ C.L. limit on $\Delta m_s$ corresponds to the value $\Delta m_s^{lim}$ for 
which ${\cal A} + 1.645~\sigma_{\cal A}$~=~1, whereas the sensitivity $\Delta m_s^{sens}$
corresponds to $ 1.645~\sigma_{\cal A}$~=~1. It is the value of $\dms$ at which it is expected
that the 95$\%$ limit will be set, if ${\cal A}$~=~0.  
In addition, the final likelihood distribution can be retrieved from
the combined amplitudes corresponding to all analyses as explained in the following.

The information coming from oscillation amplitude measurements, which
gives constraints on possible values of $\dms$, was 
included \cite{ref:pprs,ref:parodietal,ref:taipei,ref:Bphysb,ref:plasz}, 
up to now, using the distribution~\footnote{ The
dependence of $\cal A$ and $\sigma_{\cal A}$ on $\dms$ 
is always implicitly assumed.}
\begin{equation}
{\rm F}(\dms)=~
\exp {-\frac{1}{2}\left ( \frac{{\cal A}-1}{\sigma_{\cal A}} \right )^2}.
\label{dms_wrong}
\end{equation}

Recently it has been proposed to use the log-likelihood function
$\Delta \log{\cal L}^{\infty}(\dms)$ referenced to its value obtained 
for $\dms=\infty$~\cite{ref:checchia1}. 
Similar considerations, developed in a different
context, have been detailed in 
\cite{guru}.
The log-likelihood values can be easily deduced
from $\cal A$ and $\sigma_{\cal A}$ using the expressions given in~\cite{moser}
\renewcommand\arraystretch{1.1}
\begin{eqnarray}
\Delta \log{\cal L}^{\infty}(\dms) &
= \frac{\textstyle 1}{\textstyle 2}\,\left[ \left(\frac{\textstyle {\cal A}-
1 }{\textstyle \sigma_{\cal A} }\right)^2-
 \left(\frac{\textstyle {\cal A}}{\textstyle\sigma_{\cal A}}\right)^2 \right]
& = \left (\frac{\textstyle 1}{\textstyle 2}
-{\cal A}\right){\textstyle 1\over 
   \textstyle \sigma_{\cal A}^2} 
\ , 
\label{dms_right1}\\
&\Delta \log{\cal L}^{\infty}({\dms})_{mix}   &= -\frac{1}{2}\frac{1}{\sigma_{\cal A}^2} 
\label{dms_right2} \ ,  \\
&\Delta \log{\cal L}^{\infty}({\dms})_{nomix} &=  \frac{1}{2}\frac{1}{\sigma_{\cal A}^2 }
\ . \label{dms_right3}
\end{eqnarray} 
\renewcommand\arraystretch{1.0}
The last two equations give the average log-likelihood value when
$\dms$ corresponds to the true oscillation frequency
({\it mixing} case) and when $\dms$ is far from the oscillation frequency
($|\dms-\dms^{true}| \gg \Gamma/2$, {\it no-mixing} case).
$\Gamma$ is the full width at half maximum of the amplitude distribution in case of a signal; typically 
$\Gamma \simeq {1}/{\tau_{\Bs}}$.
In the following it is shown that:
\begin{itemize}
\item Equation~(\ref{dms_wrong}) does not represent the optimal
      way to include the $\Delta m_s$ information.\\
      In particular it is incorrect in case
      of a measurement;
\item Equation~(\ref{dms_right1}) allows to define the likelihood ratio R
      \begin{equation}
        {\rm R}(\dms) = e^{\textstyle {-\Delta \log {\cal L}^{\infty}(\dms)}} =
	\frac{{\cal L}(\dms)}{{\cal L}(\dms=\infty)} \ .
      \label{R_eq}
      \end{equation}
The function R corresponds to the ratio of probability densities for different
$\Delta m_s$ values.
The absolute $\dms$ probability density, instead,
remains undefined because ${\cal L}(\dms)$ stays constant 
for $\dms$ values much larger than the sensitivity;
in this region R is equal to unity.
\item a reasonable procedure to extrapolate $\rm R$ in regions where the direct measurements
      of the amplitudes are not available can be established.
\end{itemize}
\subsection{Comparison between the new and the old methods}
The main concern, in the previous approach, was that the sign of the deviation
with respect to the value ${\cal A} = 1$ was not used, whereas it is expected
that an evidence for a signal would manifest itself by giving an amplitude value
which is simultaneously compatible with ${\cal A} = 1$ and
incompatible with ${\cal A} = 0$. 
\par
\begin{figure}[htb!]
\begin{center}
\epsfig{file=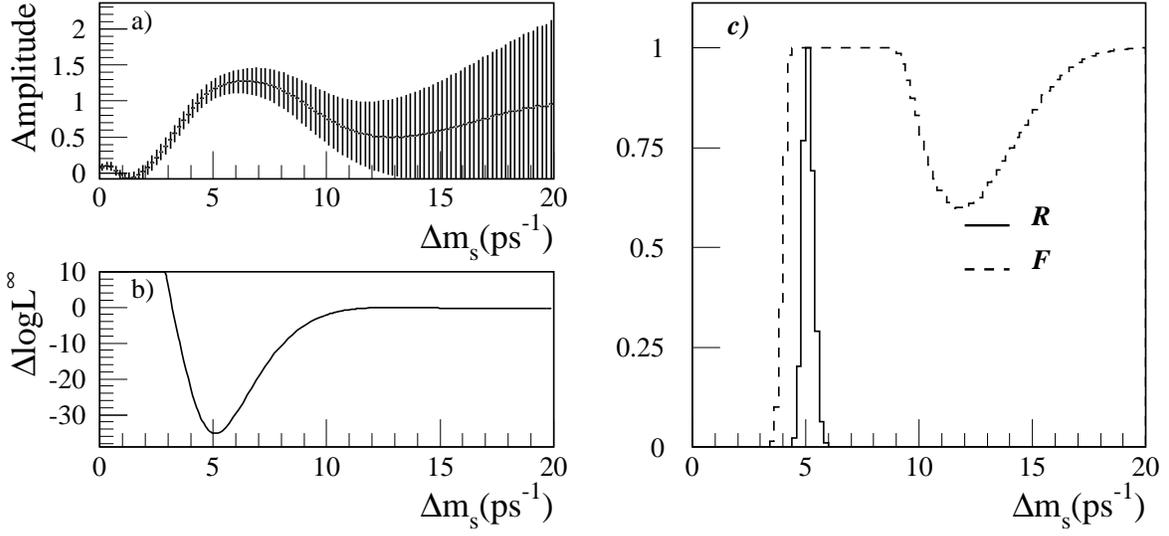,width=16cm}
\end{center}
\caption{\small { Amplitude analysis with $\dms$ generated at 5~$\mbox{ps}^{-1}$:
         a) amplitude spectrum, b) $\Delta \log {\cal L}^{\infty}(\dms)$,
         c) comparison between R and the function F used in the previous method.
         The maxima of the two functions have been rescaled to unity in order to ease the
         comparison}}
\label{fig:amp_toy}
\end{figure}

\begin{figure}[htb!]
\begin{center}
\epsfig{file=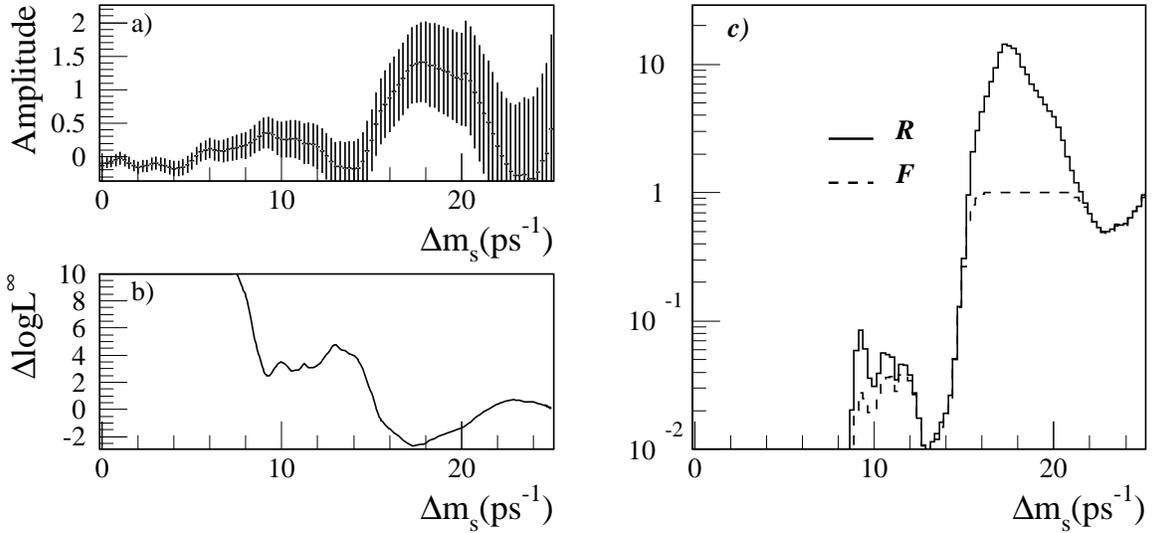,width=16cm}
\end{center}
\caption{\small{ World average amplitude analysis:
         a) amplitude spectrum, b) $\Delta \log {\cal L}^{\infty}(\dms)$,
         c) comparison between ${\rm R}$ and the function used in the previous method.}}
\label{fig:amp_avg}
\end{figure}

  The new method, at variance, includes
the relative weight of the two hypotheses.
Equations~(\ref{dms_right1}) and~(\ref{R_eq}) show
that $\rm R$ has a clear interpretation as ratio between two
cases: mixing (${\cal A} = 1$) or no mixing (${\cal A} = 0$).
The properties of the log-likelihood ratio, R, and its application in Higgs boson searches have been 
introduced in~\cite{ref:RDago}.\\
Another problem of the previous approach is that the sign of the deviation
of the amplitude with respect to unity is not considered: this implies
that a lower probability is attributed to $\dms$ values with ${\cal A} > 1$ with respect 
to $\dms$ values having ${\cal A} = 1$. Since this behaviour is clearly undesired,
in the old method the amplitudes larger than unity have been set to unity
(as it has been implicitly done in \cite{ref:pprs}-\cite{ref:taipei}).\\
The difference between the old and the new method are illustrated in two cases:
\begin{itemize}
\item a single analysis having the sensitivity at 15~ps$^{-1}$
 (see next paragraph and Figure \ref{fig:amp_toy}) simulating a signal
at $\dms=5$ps$^{-1}$ (this value has been chosen well below
the sensitivity in order to amplify the effect).
\item the world average analysis (see Figure \ref{fig:amp_avg})
\end{itemize}

In Figure~\ref{fig:amp_toy}-a and Figure~\ref{fig:amp_toy}-b,
respectively, the  amplitude
spectrum ${\cal A}$
and the corresponding distribution for $\Delta {\cal L}^{\infty}(\dms)$
are  given for  the simulated case,
in which  there is  
a signal at $\Delta m_s =5~ {\rm ps}^{-1}$.
Figure~\ref{fig:amp_toy}-c   shows that
there is  a marked  difference between the behaviour of ${\rm R}$ and
${\rm F}$. While the function $\rm R$ shows, by construction, a maximum corresponding
to the fitted value ($\dms=(5.10\pm 0.25)~\mbox{ps}^{-1}$), the function $\rm F$
is not able to spot the maximum, in particular it attributes large
probability to any $\dms$ value with ${\cal A} \sim  1$
regardless its error.\\
For the world average analysis, Figure~\ref{fig:amp_avg} shows that the 
agreement between $\rm R$ and $\rm F$ is acceptable only when ${\cal A} \sim 0$.

\subsection{How to treat \boldmath$\dms$ regions without amplitude measurements?}
\par
\begin{figure}
\begin{center}
\epsfig{file=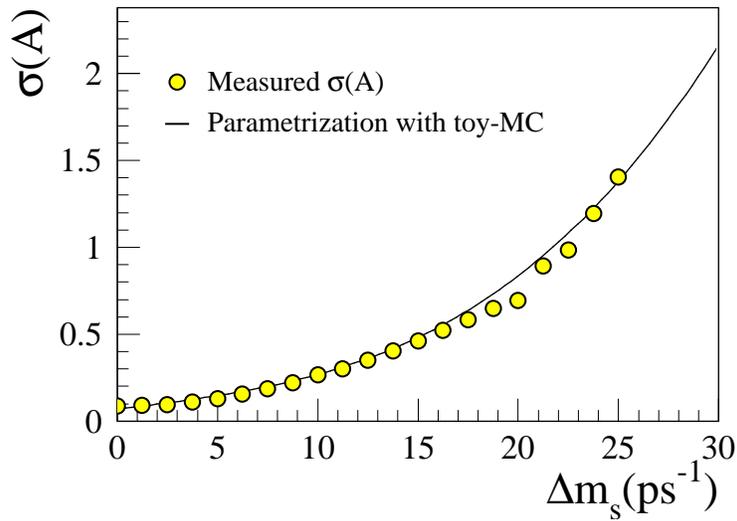,width=10cm}
\end{center}
\caption{\small{ Comparison between the error distribution computed with a toy-MC
            (solid line) and the measured amplitude errors (circles).
}}
\label{extra_dms}
\end{figure}

 The procedure used to continue $\rm R$
 beyond the last measured amplitude value ($\dms^{last}$) is based on the continuation
of the amplitude spectrum from which $\rm R$ is deduced:
\begin{itemize}
  \item continuation of $\sigma(\cal A)$:
        the behaviour of $\sigma(\cal A)$ for $\dms>\dms^{lim}$ can be reproduced
        by tuning the parameters of a fast simulation
        (toy-MC).
        The method used here is similar to the one presented in
       ~\cite{ref:boix}.
        The errors on the amplitude can be written as:
        \begin{eqnarray*}
          \sigma^{-1}({\cal A})= \sqrt{N}\, \eta_{{B_s}}
	  \, (2\epsilon-1) \, W(\sigma_L,\sigma_P,\dms)
        \end{eqnarray*}
        where $N$ is the total number of events, $\eta_{{B_s}}$ the purity
	 of the
        sample in ${B_s}$
        decays, $\epsilon$ the purity of the tagging at the decay time,
        $\sigma_L$ is the ${B_s}$ flight length uncertainty and
        $\sigma_P$ the relative uncertainty on its momentum.
        The parameters $\sigma_L$, $\sigma_P$ and the global factor that
        multiply $W$ have been obtained by adjusting the simulated 
        error distribution on the one measured with real events.
        \par
        Figure~\ref{extra_dms} shows the agreement between the toy-MC calculation
        and real data for $\dms<\dms^{last}$ (the upper bound on $\dms$
        of the amplitude plot for which there are measurements) and the extrapolation to higher values. 
        It has been verified (with an independent toy-MC) that the extrapolated
        errors approximate well the simulated error distribution.
  \item continuation of $\cal A$: this part is more critical than the previous
        one. In particular it is more sensitive to the real amplitude spectrum.
        Nevertheless if $\dms^{sens} << \dms^{last}$, the significance $S$
        ($S={\cal A}/\sigma_{\cal A}$) is approximately constant.
        
\end{itemize}
In the following the amplitudes at $\dms>25~\mbox{ps}^{-1}$ 
(which corresponds to $\dms^{last}$ for the combined world average plot)
are deduced using the extrapolation of the amplitude errors
and fixing the significance at the last measured value.
As a consequence, for $\dms$ values larger than $\dms^{last}$, R is
fixed to the last measured point.
Although this procedure is reasonable, it should be stressed that
it is very desirable to have all the amplitudes (with errors) up to the
$\Delta m_s$ value where  $\rm R$ approaches its plateau.
\subsection{The likelihood function $\rm R$ used in the analysis}
The present world average on $\dms$ shows an ``hint'' of  signal at
$\dms \sim$ 17.5~ps$^{-1}$ (see  Figure~\ref{fig:amp_avg}-b). Its significance (about 2.2 $\sigma$)
is not sufficient to claim evidence for an observation of the ${B}^0_s-\bar{{B}}^0_s$ oscillation. 
This effect is in fact larger than the average value, of $\sim 1.2 \,\sigma$, which is expected 
in case of a real signal in that region. \\
Whatever conclusion can be drawn from these data, it should be stressed that:
\begin{itemize}
\item the likelihood function $\rm R$ deduced from data contains
the present knowledge 
      on $\dms$ and it is the optimal weight function to be used in the fit;
\item the use of $\rm R$ does not imply any assumption 
either on the evidence or on the presence of a signal.
It only translates the experimental fact that the log-likelihood function 
has  a minimum at 17.5~ps$^{-1}$ with a $2.2~\sigma$ significance.
\end{itemize}
%This point  is clear if we compare   the observed  $\dms$ likelihood
%with that expected  in case of  measurement. 
%{Rome begin could you please translate the following sentences in any
%indoeuropean language ? }
%When a certain quantity is measured to be
%$(x\pm\sigma)$ no one would be shocked if a more precise measurement will turn out to
%be 2.2 $\sigma$ away from the present value and, at the same time, this would not prevent
%anybody  to use the present available information.\\
%The $\dms$ likelihood is simply downweighting all the values between $\dms=14.0~ps^{-1}$
%and $\dms=\infty$ with respect to $\dms=18.0~ps^{-1}$
%%by a factor that is, at worst, $exp(-(2.2)^2/2)$, the same factor
%we would apply to a value that is 2.2 $\sigma$ away from the present measured 
%value. {\bf Rome end}
%{\bf Rome begin I have eliminated a sentence Rome end}
%Future $\dms$ measurements will certainly improve our knowledge, but this should not
%prevent us from using the actual data.

\begin{figure}
\begin{center}
\epsfig{file=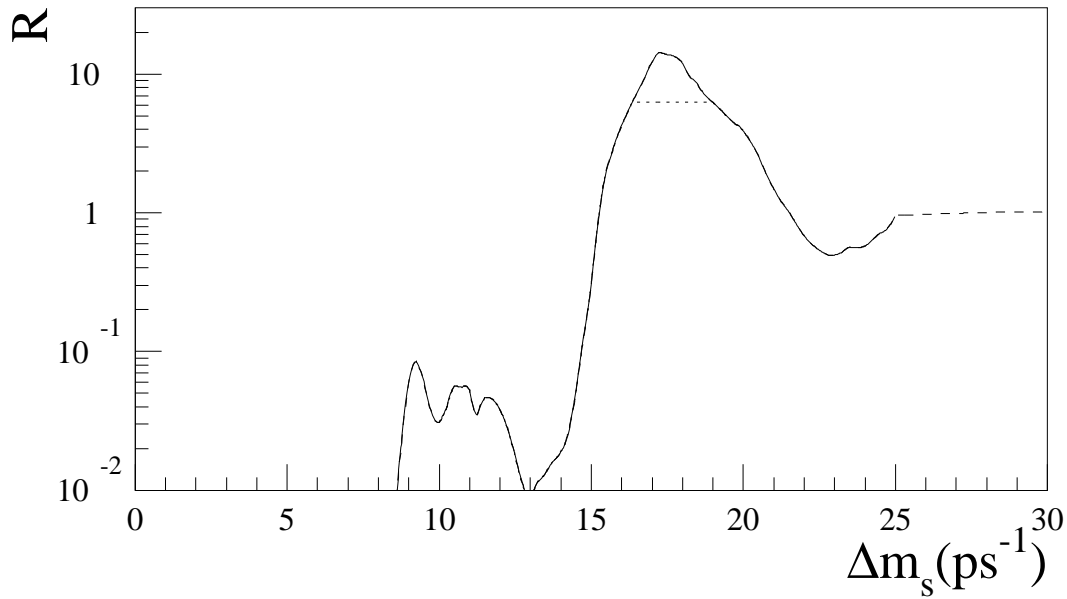,width=14cm}
\end{center}
\caption{\small{ Comparison between different weight functions used in the fit.
The curve determined by the continuous and dashed lines corresponds to the
likelihood ratio obtained with, respectively, the measured amplitudes and the
extrapolation to large values of $\dms$. The dotted curve is obtained by truncating 
the previous distribution once the amplitude becomes larger than one.}}
\label{R_dms}
\end{figure}

Although we believe that the use of $\rm R$ is, indeed, the best method 
to include the information  on $\dms$, we briefly mention some  alternative approaches. The main difference 
is that, with these methods, a fraction of the available experimental information is lost.
\begin{enumerate}
  \item 95\% C.L. limit: this method uses a stepwise function 
starting at $\dms^{lim}$. We mention it  only for 
completeness since the  choice of the C.L. is arbitrary.
\item Min($\rm R$,1): in this approach it is considered
that no value of $\dms$ has to be preferred   
with respect to $\dms=\infty$ 
(only exclusion is possible).
It was used in our preliminary analyses \cite{conferences} 
and consists in setting ${\cal A}=0.5$ for all amplitudes having larger values (see equation~(\ref{dms_right1}).
Since there is no reason to throw away the information
contained in  amplitudes with values ranging between 0.5 and 1, this method has been
abandoned.
\item Min[${\rm R}(A=1)$,$\rm R$]: in this method one
tries to avoid biases induced by ``lucky"  fluctuations. The strategy is then 
to limit the likelihood ratio to the  
value obtained when the  amplitude  reaches ${\cal A}=1$ for the first time.
The effect is to  flatten the ratio R around its maximum.
\end{enumerate}
\par
Two different definitions of $\rm R$ have then been used as weight function in the 
fitting procedure:
\begin{itemize}
  \item the complete $\rm R$ distribution deduced from data and from the toy-MC;
  \item the function Min[${\rm R}(A=1)$,$\rm R$]. 
\end{itemize}
They are shown in Figure~\ref{R_dms} and the second choice has been considered as 
a ``pessimistic'' approach to evaluate the induced variation on the fitted quantities in Section \ref{sec:stability}.

\section{Results and discussion}
\label{sec:results}
In this Section we give the results for the quantities of interest.
%$\rhobar$, $\etabar$, sin$(2\beta)$, sin$(2\alpha)$ and $\gamma$  
Values of the hadronic parameters, which can be determined
quite accurately by assuming the validity of the Standard Model, are also extracted.
The central value is always given using the average, and the error corresponds to 
the standard deviation. For asymmetric p.d.f. we also give the median and the error
corresponds to regions containing 34\% probability on each side of the median.

\subsection{Results obtained with all measurements}

%%%%%%%%%%%%%%%%%%%%%%%%%%%%%%%%%%%%%%%%%%%%%%%%%%%%%%%%%%%
\begin{figure}
\begin{center}
{\epsfig{figure=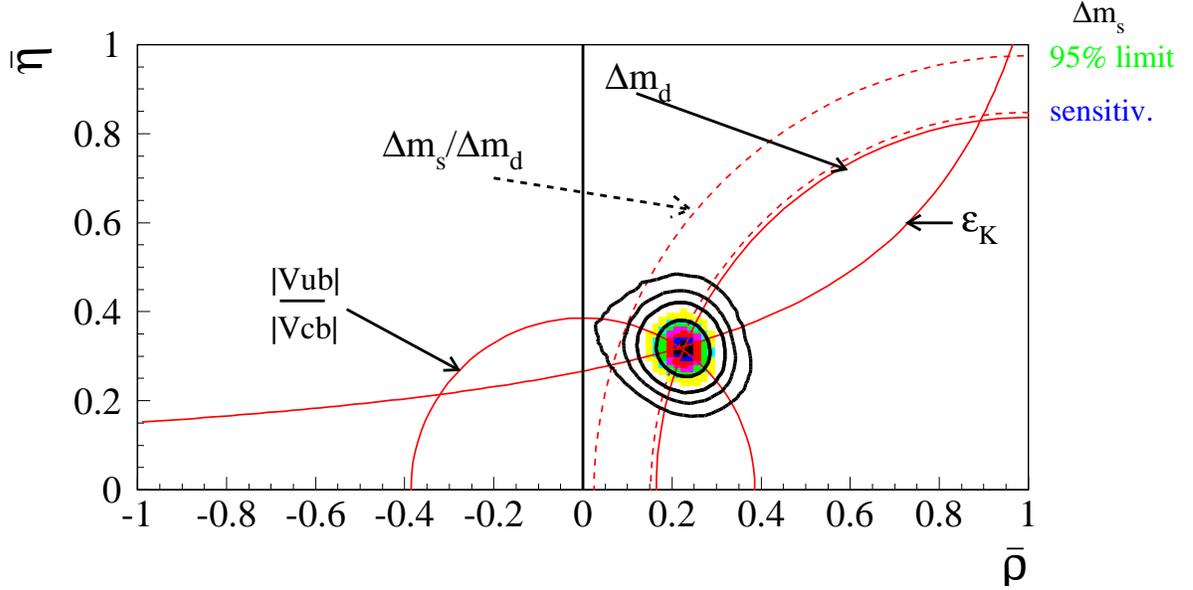,bbllx=30pt,bburx=503pt,bblly=1pt,bbury=270pt,height=8cm}}
\caption{ \small{Allowed regions for $\rhobar$ and 
$\etabar$ using the parameters listed in Table~\ref{tab:1}.
The contours at 68\%, 95\%, 99\% and 99.9\% probability are shown. The full lines 
correspond to the central values of the constraints given by
the measurements of  $\left | V_{ub} \right |/\left | V_{cb} 
\right |$, $\epsilonk$ and $\Delta m_d$.
The two dotted curves correspond, from left to right respectively, to the 95\% upper limit and 
to the value of the sensitivity obtained from the experimental study of ${B}^0_s-\bar{{B}}^0_s$ oscillations.}}
\label{fig:rhoeta}
\end{center}
\end{figure}

%%%%%%%%%%%%%%%%%%%%%%%%%%%%%%%%%%%%%%%%%%%%%%%%%%%%%%%%%%%
\begin{figure}
\begin{center}
{\epsfig{figure=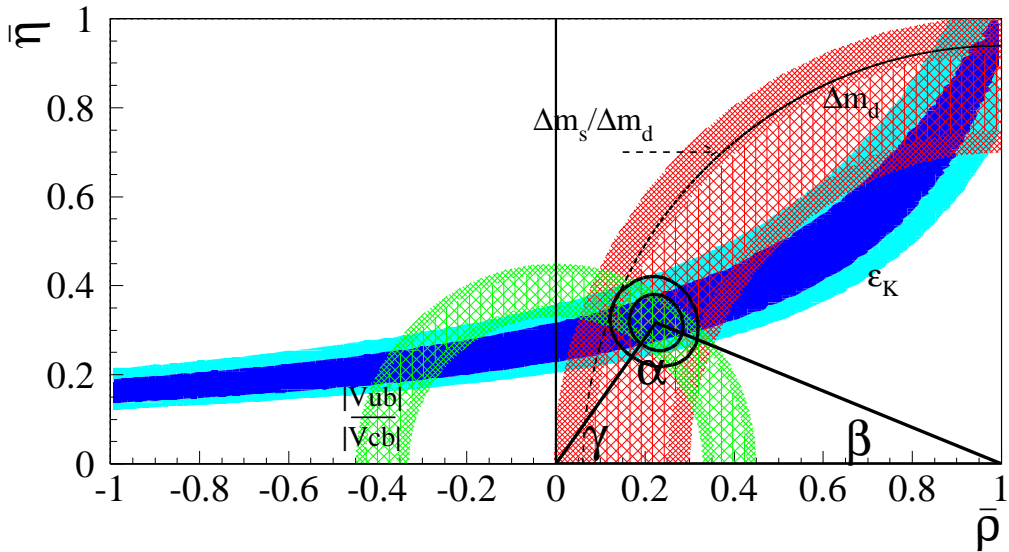,
bbllx=30pt,bburx=503pt,bblly=1pt,bbury=270pt,height=8cm}}
\caption{ \small{The allowed regions for $\rhobar$ and 
$\etabar$ (contours at 68\%, 95\%) are compared with the uncertainty bands (at 68\% and 95\% probabilities)
for $\left | V_{ub} \right |/\left | V_{cb} \right |$, $\epsilonk$, $\Delta m_d$ and the 
limit on $\Delta m_s/\Delta m_d$ (dotted curve).}}
\label{fig:bands}
\end{center}
\end{figure}
%%%%%%%%%%%%%%%%%%%%%%%%%%%%%%%%%%%%%%%%%%%%%%%%%%%%%%%%%%%

\begin{figure}
\begin{center}
{\epsfig{figure=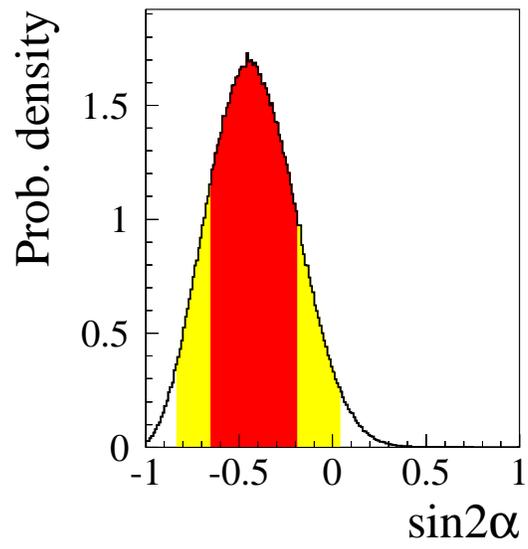,bbllx=1pt,bburx=250pt,bblly=1pt,bbury=250pt,height=7.5cm}}\\
{\epsfig{figure=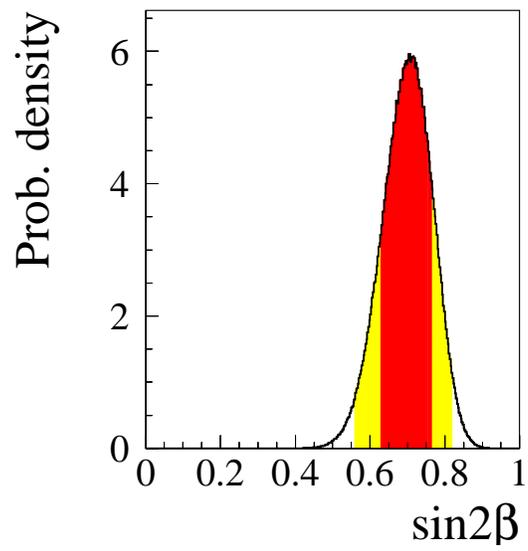,bbllx=1pt,bburx=250pt,bblly=1pt,bbury=250pt,height=7.5cm}}\\
{\epsfig{figure=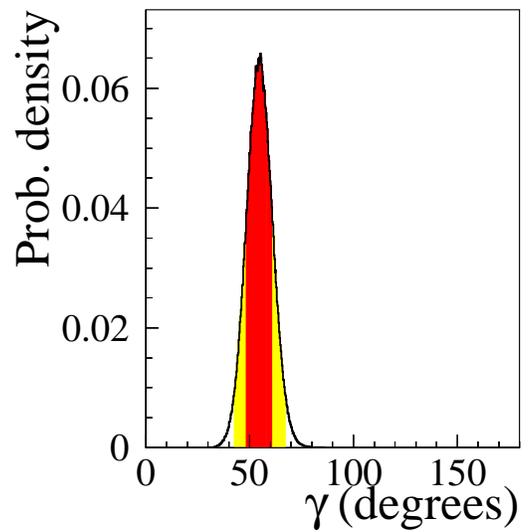,bbllx=1pt,bburx=250pt,bblly=1pt,bbury=250pt,height=7.5cm}}
\caption{\small {The p.d.f. for $\sin(2\alpha)$, $\sin(2\beta)$ and $\gamma$ . The red (darker) 
and the yellow (clearer)  zones correspond respectively to 68\% and 95\% of the normalised area. }}
\label{fig:alphabetagamma}
\end{center}
\end{figure}

The region in the ($\rhobar,~\etabar)$ plane selected by the measurements
of $\epsilonk$, $\vubsvcb$, $\dmd$ and from the information 
on $\dms$ (using the R function of Figure \ref{R_dms}) 
is given in Figure~\ref{fig:rhoeta}.  In Figure~\ref{fig:bands}
the uncertainty bands for the quantities, obtained using 
Equations~(\ref{eq:C_vubovcb})--(\ref{eq:epskdef}), are presented.
Each band, corresponding to only  one of  the constraints,  
contains 68\% and 95\%  of the events obtained by varying  the input
parameters. This comparison illustrates the consistency of the
different constraints provided by the Standard Model.
In the present studies, two statistically equivalent procedures have been used.
They differ in the way  values for $\bar\rho$ and $\bar\eta$ have been extracted 
and in the inclusion of QCD corrections to $\epsilon_K$ and to ${B}^0-\bar{{B}}^0$
mixing, which are either computed~\cite{ref:roma} or  taken as independent 
inputs~\cite{ref:parodietal}.
The measured values of the  two parameters are
\begin{equation}
\rhobar=0.224 \pm 0.038  ,~\etabar=0.317 \pm 0.040  
\label{eq:eta1}
\end{equation}
and 
\begin{equation}
\rhobar=0.221 \pm 0.037  ,~\etabar=0.315 \pm 0.039 \ ,
\end{equation}
using the methods of ref.~\cite{ref:parodietal}  and \cite{ref:roma} 
respectively.
The two quantities are practically uncorrelated (correlation coefficient of -5\%),
as it can be seen from the contour plot of Figure~\ref{fig:rhoeta}.
%The correlation between $\rhobar$ and $\etabar$ is $corr(\rhobar,\etabar)$= -0.05.\\
Fitted values for the angles of the unitarity triangle have been obtained also
\begin{equation}
\sin(2\,\beta)= 0.698 \pm 0.066\, ,\quad 
\sin(2\,\alpha)= -0.42 \pm  0.23\, , \quad \gamma=(54.8 \pm 6.2)^{\circ}
\quad \cite{ref:parodietal}
\label{eq:eta2}
\end{equation}
(the quoted error is symmetric, in spite of the small asymmetry of the distributions
shown in Figure~\ref{fig:alphabetagamma})
and 
\begin{equation}
\sin(2\beta)= 0.692 \pm 0.065\, ,\quad \sin(2\alpha)= -0.43 \pm  0.21\, , 
\quad \gamma=(54.9 \pm 5.7)^{\circ} \quad \cite{ref:roma}.
\end{equation}
respectively. 

The results in the two approaches are in agreement and, in the following, quoted values and figures 
are those obtained using the approach of~\cite{ref:parodietal}. 
In Figure~\ref{fig:alphabetagamma}, the p.d.f. for the angles of the unitarity triangle are given.
%%%%%%%%%%%%%%%%%%%%%%%%%%%%%%%%%%%%%%%%%%%%%%%%%%%%%%%%%%%
%%%%%%%%%%%%%%%%%%%%%%%%%%%%%%%%%%%%%%%%%%%%%%%%%%%%%%%%%%%
Few comments can be made:
\begin{itemize}
\item $\sin(2\beta)$ is determined quite accurately. This value has
to be compared with recent measurements of this quantity using $J/\psi K_S$
events from LEP~\cite{lepsin2b}, CDF~\cite{cdfsin2b}, BaBar~\cite{ref:babarsin2b} 
and BELLE~\cite{ref:bellesin2b}. The accuracies of these measurements are completely 
dominated by the statistical errors and their average is $\sin(2\beta) = 0.49 \pm 0.16$.

\item the angle $\gamma$ is known within an accuracy of about 10\%. It has to be stressed that,
with present measurements, the probability that
$\gamma$ is greater than 90$^{\circ}$ is only 0.03\%.
Without including the information from $\dms$, it is found that $\gamma$ has
4\% probability to be larger than 90$^{\circ}$.
The central value for the angle $\gamma$ is much smaller than that obtained
in recent fits of rare $B$-meson two-body  decays~\cite{ref:cleorare}.
It remains to be seen to which extent  the results of~\cite{ref:cleorare}
are affected by the  model dependence in the
theoretical description of two-body decays.
In this respect, it would be interesting to examine
under which conditions  these decays can be  described
 by the same value of $\gamma$ as found in the present study.
An exploratory work in this direction can be found in \cite{zhou}.
\end{itemize}

%Finally, for completeness, we also give the value for $\mbox{Im}\,\lambda_t$
%\footnote{We recall here that $\mbox{Im}\,\lambda_t = \lambda \Vcb^2 \etabar$.}
%\begin{eqnarray}
%\mbox{Im}\,\lambda_t = (1.19 \pm 0.12)~10^{-4}
%\end{eqnarray} 

\subsection{The CKM triangle from \boldmath$b$-physics alone}
\begin{figure}
\begin{center}
{\epsfig{figure=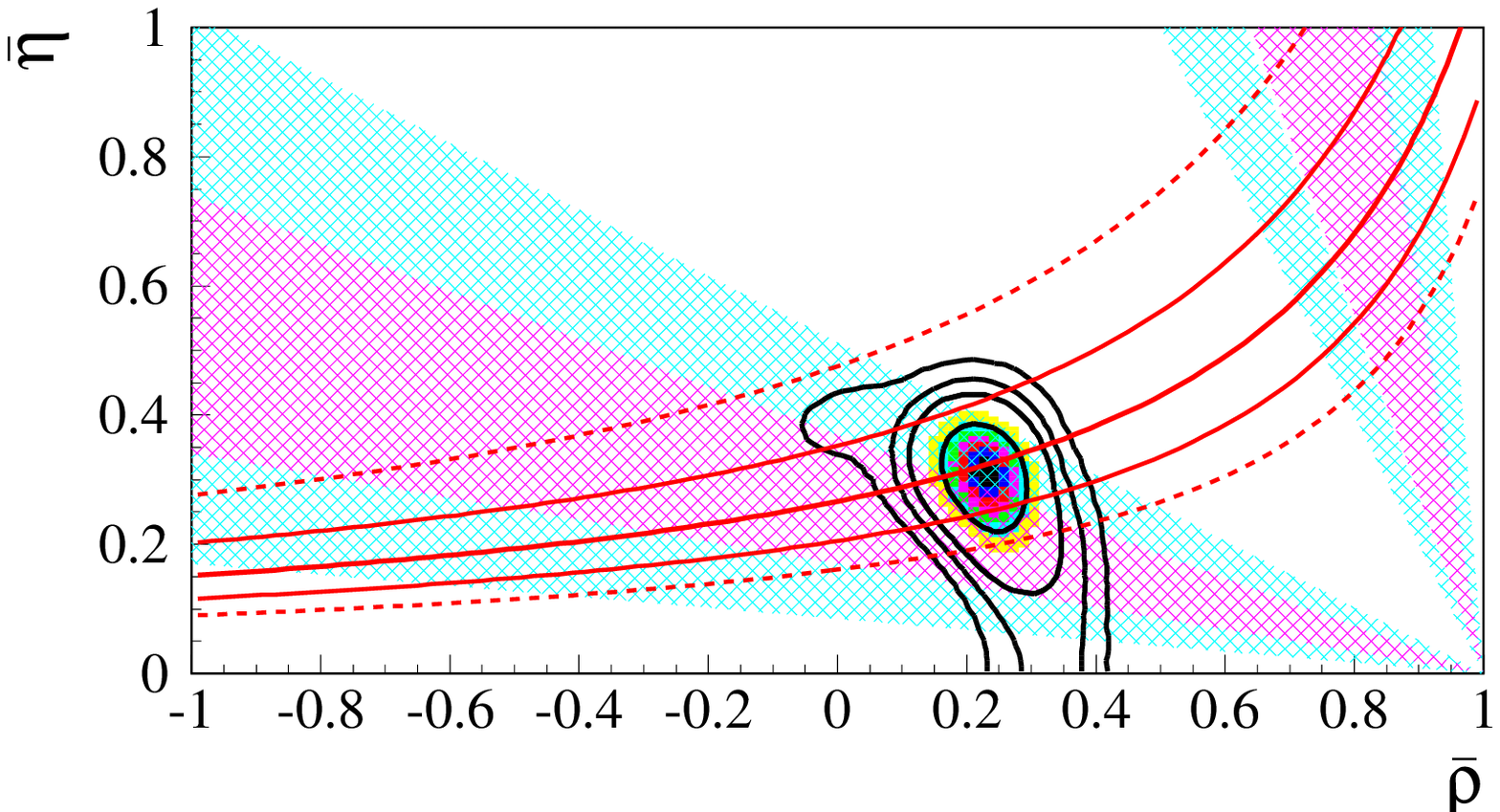,bbllx=30pt,bburx=503pt,bblly=1pt,bbury=270pt,height=8cm}}
\caption{\small {The allowed regions (at 68\%, 95\%, 99\% and 99.9\% probability) for $\rhobar$ and
$\etabar$ using  the constraints given by the measurements of  $\left | V_{ub} \right |/\left | V_{cb}
\right |$,  $\Delta m_d$ and  $\Delta m_s$. The constraint due to $\epsilonk$ is not included. The regions
(at 68\% and 95\% probability) selected by the measurements of $\epsilonk$ (continuous (1$\sigma$) and dotted (2$\sigma$) curves)
and $\sin(2\,\beta)$ (darker (1$\sigma$) and clearer (2$\sigma$) zones) are shown. 
For $\sin(2\,\beta)$ the two solutions are displayed.}}
\label{fig:noepk}
\end{center}
\end{figure}

As four constraints are used to determine the values of two parameters,
it is possible to relax, in turn,  one (or more)  of these constraints,
still obtaining significant confidence intervals.
An interesting exercise consists in  removing
the theoretical constraint for  $\hat B_K$ in  the measurement of $\epsilonk$
(\cite{ref:checchia}-\cite{ref:noepsk}).
 The corresponding selected region in the ($\rhobar,~\etabar)$ plane
is shown
in  Figure~\ref{fig:noepk}, where the region selected by the
measurement of $\epsilonk$ alone is also drawn.
This comparison shows that the Standard Model picture of CP violation in the $K$ system and
of $B$ decays and oscillations are consistent.
In the same figure, we also compare the allowed regions in the ($\rhobar,~\etabar)$ plane
with those selected by the measurement of $\sin(2\,\beta)$ using $J/\psi K_S$ events.

%%%%%%%%%%%%%%%%%%%%%%%%%%%%%%%%%%%%%%%%%%%%%%%%%%%%%%%%%%%
%%%%%%%%%%%%%%%%%%%%%%%%%%%%%%%%%%%%%%%%%%%%%%%%%%%%%%%%%%%

Using constraints from $b$-physics alone the following results are obtained
%\begin{equation}
%\etabar= 0.325 \pm 0.062\,, \quad \sin(2\beta)= 0.694 \pm 0.097\, ,
%\label{eq:eta3}
%\end{equation}
%or, using the median of the distribution:
\begin{eqnarray}
\etabar     = 0.302^{+0.052}_{-0.061}  \quad ;&  \quad[0.145-0.400] \quad \rm{at}~95\% \quad ;& > 0.08 ~\rm{at}~ 99\% \nonumber \\
\sin(2\,\beta) = 0.678^{+0.078}_{-0.101}  \quad ;&  [0.392-0.818] \quad \rm{at}~95\%  &        
\label{eq:eta32}  
\end{eqnarray}

\begin{figure}
\begin{center}
{\epsfig{figure=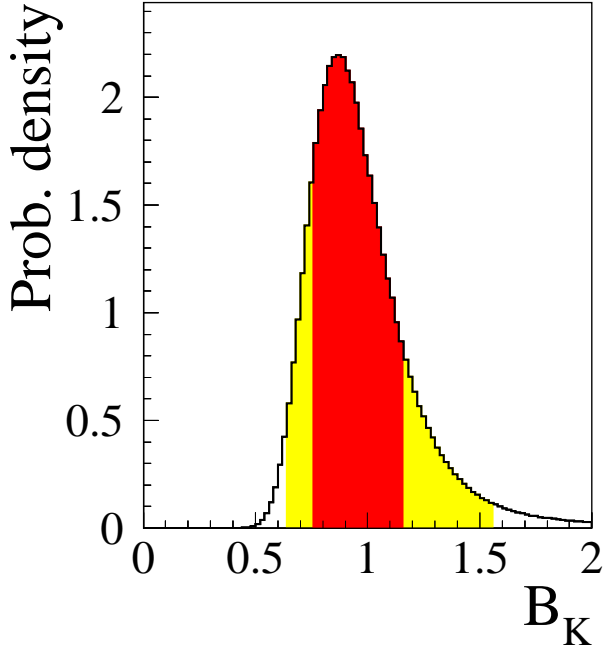,bbllx=1pt,bburx=305pt,bblly=1pt,bbury=250pt,height=9cm}}
\caption{ \small{The p.d.f. for  $\hat \BK^{\rm b-phys}$ .}}
\label{fig:bk}
\end{center}
\end{figure}

\noindent (in terms of average and standard deviation the results are $\etabar$ = 0.296 $\pm$ 0.063 and 
$\sin(2\,\beta)$ = 0.663 $\pm$ 0.109). \\
Another way for illustrating the agreement between $K$ and $B$ measurements consists
in comparing the values of the $\hat \BK$ parameter obtained in lattice
QCD calculations with the value extracted  from
Equation~(\ref{eq:epskdef}), using the
values of $\rhobar$ and $\etabar$ selected by  $b$-physics alone
%\begin{equation}
%\hat \BK^{\rm b-phys}=0.93 \pm 0.22\, ,
%\label{eq:bkb}
%\end{equation}
%or, using the median of the distribution
\begin{eqnarray}
\hat \BK^{\rm b-phys}=0.90^{+0.30}_{-0.14}  ,&  0.64 \le \hat \BK^{\rm b-phys} \le 1.8 ~{\rm at}~ 
 95\% ~{\rm probability}
\label{eq:bkb2}
\end{eqnarray}
\noindent (in terms of average and standard deviation the result is $\hat \BK^{\rm b-phys}$ = 0.97 $\pm$ 0.23).\\
Since $\hat \BK$ is not limited from above, for the present study, probabilities are normalised assuming $\hat \BK<$5.
The p.d.f of $\hat \BK^{\rm b-phys}$ is shown in Figure \ref{fig:bk}.
%%%%%%%%%%%%%%%%%%%%%%%%%%%%%%%%%%%%%%%%%%%%%%%%%%%%%%%%%%%
%%%%%%%%%%%%%%%%%%%%%%%%%%%%%%%%%%%%%%%%%%%%%%%%%%%%%%%%%%%

It can be noticed that the values of two theoretical parameters,
 namely $f_{B_d} \sqrt{\hat B_{B_d}}$ and $\xi$ have been 
used to obtain this result. Hopefully, when accurate
direct measurements of $\sin(2\,\beta)$ will become available,
it will be possible to remove another theoretical input (or even all of them).

\subsection{The expected value for \boldmath$\dms$}
\begin{figure}[htb!]
\begin{center}
{\epsfig{figure=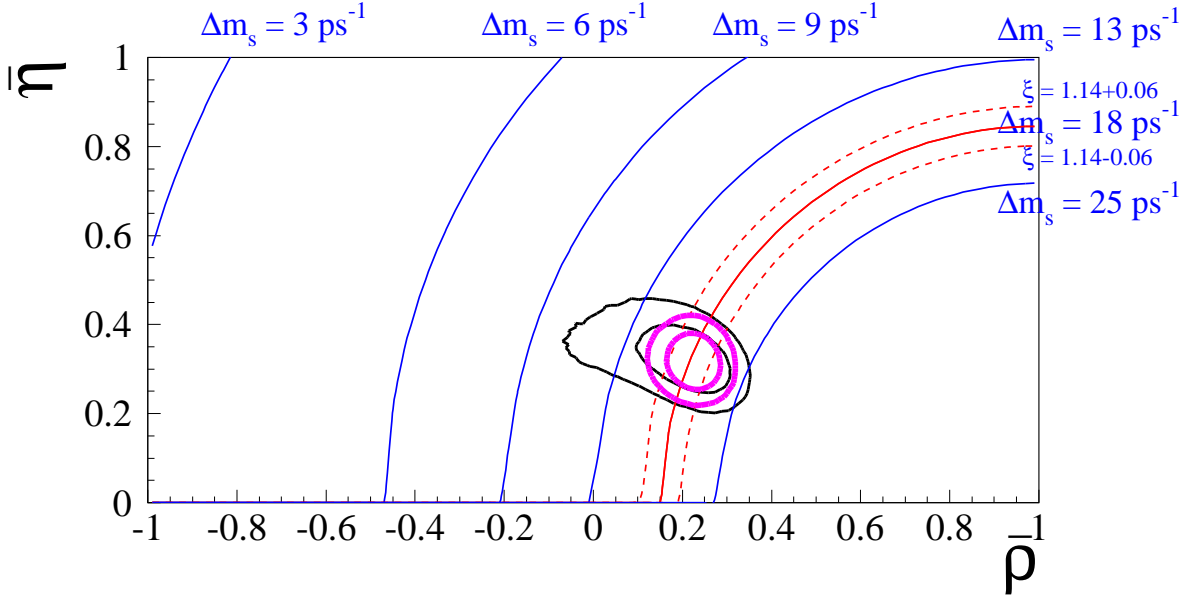,bbllx=30pt,bburx=603pt,bblly=1pt,bbury=270pt,height=8cm}}
\caption{\small {The allowed regions for $\rhobar$ and $\etabar$ using the constraints given by
the measurements of $\epsilonk$, $\left | V_{ub} \right |/\left | V_{cb} \right |$ 
and $\Delta m_d$ at 68\% and 95\% probability are shown by the thin contour lines.
Selected regions for $\rhobar$ and $\etabar$ when the constraint due to $\dms$ is included have been superimposed
using thick lines.
 The different continuous circles correspond to fixed values of $\dms$. Dashed circles, drawn on each side
of the curve corresponding to $\dms = 18.0~ps^{-1}$, indicate the effect of a variation
by $\pm 0.06$ on $\xi$.}}
\label{fig:dmsdemo}
\end{center}
\end{figure}

\begin{figure}
\begin{center}
{\epsfig{figure=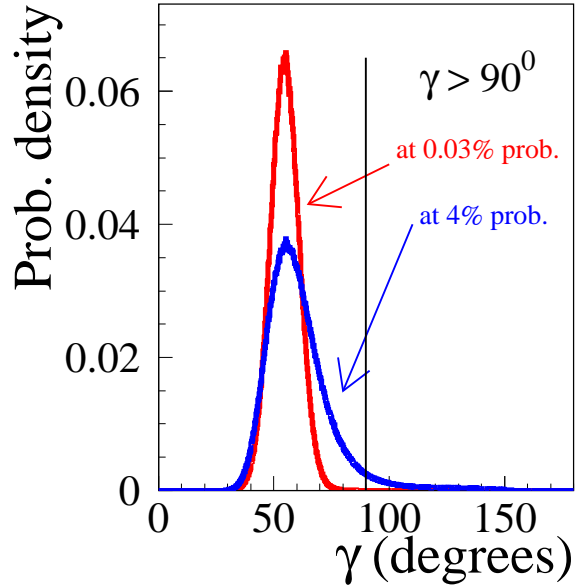,bbllx=30pt,bburx=250pt,bblly=1pt,bbury=270pt,height=9cm}}
\caption{ \small{The p.d.f. for $\gamma$ obtained with 
(red-clearer) and without (blue-darker) including the $\dms$ constraint. 
The vertical line corresponds to
$\gamma$ = 90$^{\circ}$. The probabilities of $\gamma >$ 90$^{\circ}$ 
are also given.}}
\label{fig:gammasindms}
\end{center}
\end{figure}

Figure \ref{fig:dmsdemo} shows the allowed region for $\rhobar$ and $\etabar$ 
obtained when removing the  constraint coming from the study of $\Bs$--$\Bsb$ mixing.
It illustrates the importance of the use of this information. This can be illustrated also,
from the p.d.f. of the angle $\gamma$ obtained with or without including the $\dms$ constraint 
(see Figure \ref{fig:gammasindms}). High values for $\gamma$ are excluded at high confidence
level by the experimental lower limit on $\dms$.

%%%%%%%%%%%%%%%%%%%%%%%%%%%%%%%%%%%%%%%%%%%%%%%%%%%%%%%%%%%

%%%%%%%%%%%%%%%%%%%%%%%%%%%%%%%%%%%%%%%%%%%%%%%%%%%%%%%%%%%
\begin{figure}[htb!]
\begin{center}
\begin{tabular}{cc}
\epsfig{file=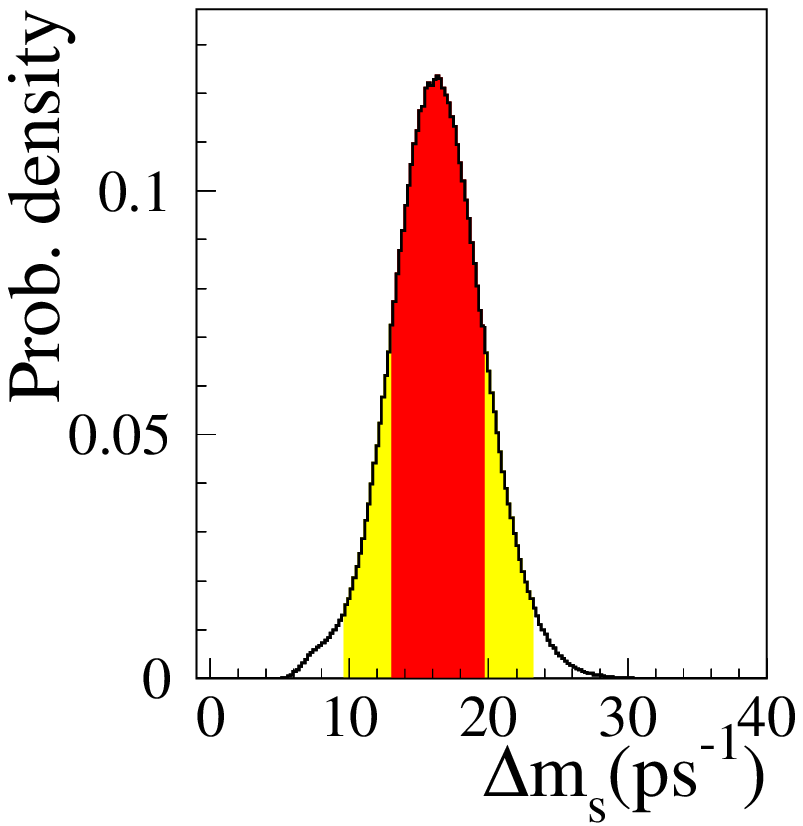, bbllx=10pt,bburx=255pt,bblly=1pt,bbury=250pt,height=8cm} & 
\epsfig{figure=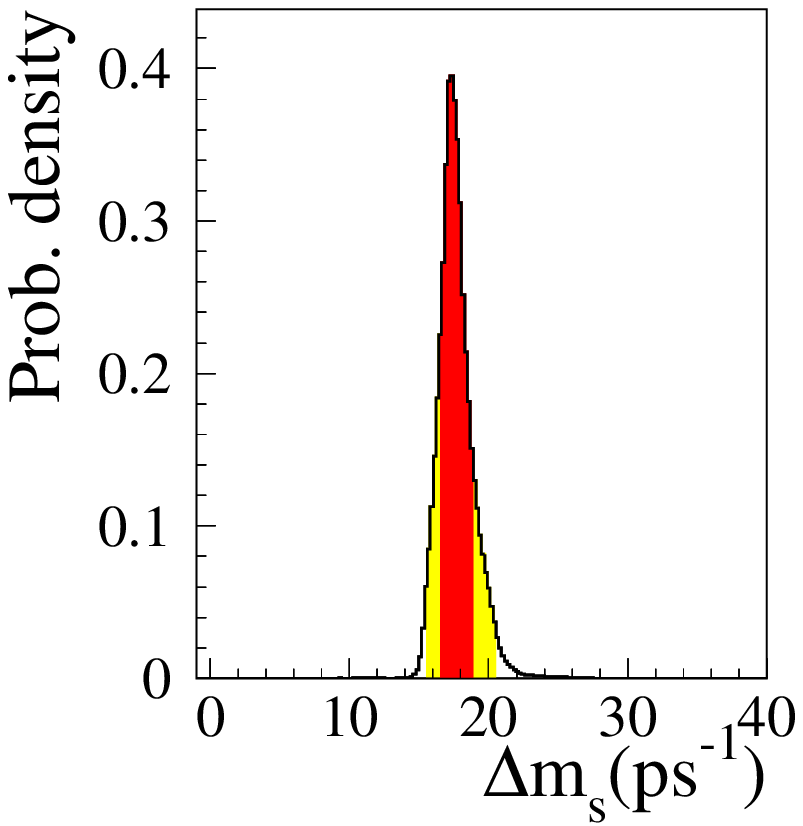, bbllx=30pt,bburx=255pt,bblly=1pt,bbury=250pt,height=8cm}
\end{tabular}
\end{center}
\caption{ \small{Probability distribution of $\Delta m_s$. The information from ${B}^0_s-\bar{{B}}^0_s$ oscillations 
 is used(not used) (distribution on the right(left).}}
\label{fig:dms}
\end{figure}
%%%%%%%%%%%%%%%%%%%%%%%%%%%%%%%%%%%%%%%%%%%%%%%%%%%%%%%%%

It is also possible to extract the probability distribution for $\dms$, which is shown in 
Figures \ref{fig:dms} (see also \cite{jaffe}). From this distribution one obtains
\begin{equation} 
\Delta m_s = (16.3 \pm 3.4) \ {\rm ps}^{-1}\, \quad \quad
9.7 \le  \Delta m_s   \le 23.2 ~{\rm ps}^{-1}\quad {\rm at}~  95\% \ {\rm probability} . \end{equation}
If the information from the ${B}^0_s-\bar{{B}}^0_s$ analyses is included, results
become
\begin{equation} 
\Delta m_s = (17.3^{+1.5}_{-0.7}) \ {\rm ps}^{-1}\, \quad \quad
15.6 \le  \Delta m_s   \le 20.5 ~{\rm ps}^{-1}\quad {\rm at}~  95\% \
{\rm probability} \end{equation}
(in terms of average and standard deviation the result is $\Delta m_s$ = 
(17.7 $\pm$ 1.3)${\rm ps}^{-1}$).\\
These values are in agreement with the recent estimate of
$\Delta m_s = 15.8 (2.3) (3.3) \ ps^{-1}$,  presented in~\cite{ape00}.

\subsection{Determination of \boldmath$\fbdsqbd$}

The value of  $\fbdsqbd$ can be obtained by removing
the theoretical constraint coming from this parameter in $\Bd$--$\Bdb$ oscillations.
%%%%%%%%%%%%%%%%%%%%%%%%%%%%%%%%%%%%%%%%%%%%%%%%%%%%%%%%%%%
\begin{figure}
\begin{center}
{\epsfig{figure=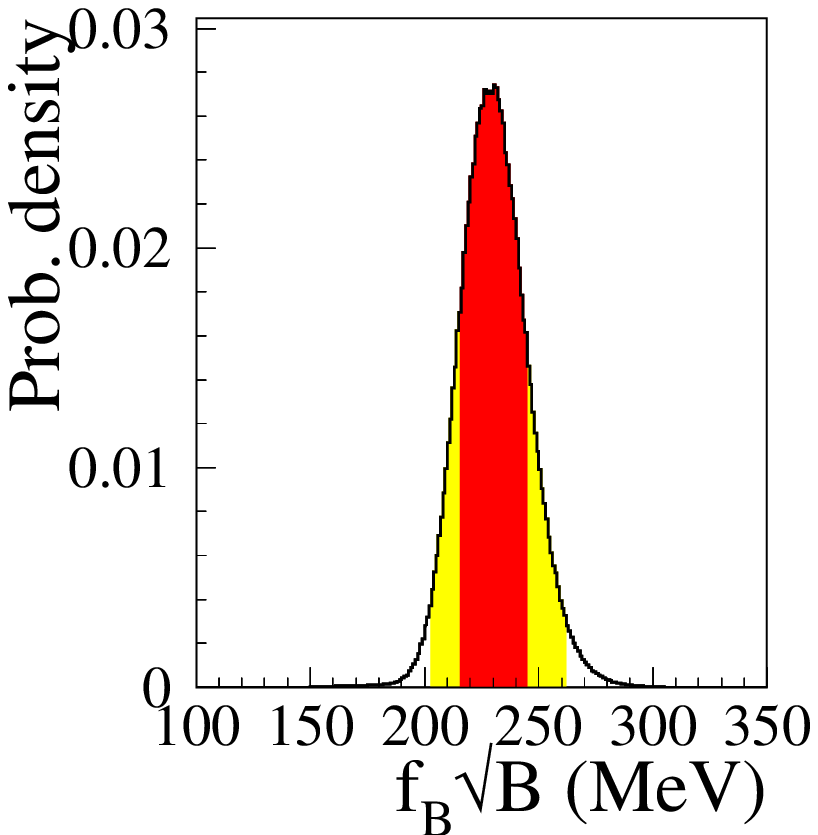,bbllx=1pt,bburx=250pt,bblly=1pt,bbury=270pt,height=9cm}}
\caption{\small {The p.d.f. for $\fbdsqbd$.}}
\label{fig:fb}
\end{center}
\end{figure}
%%%%%%%%%%%%%%%%%%%%%%%%%%%%%%%%%%%%%%%%%%%%%%%%%%%%%%%%%%%
Using the two other theoretical inputs, $\hat \BK$
and $\xi$ ,  $\fbdsqbd$ is measured with an accuracy
which is better than the current evaluation from lattice QCD, given
in Section~\ref{sec:parth}. From the p.d.f. shown in Figure~\ref{fig:fb},
\begin{equation}
\fbdsqbd = (231 \pm 15) \, \MeV \,
\end{equation}
is obtained. 
%This result is at variance with frequent statements~\cite{ref:Bphysb}--\cite{ref:stone} claiming that
%the high accuracy, obtained on $\etabar$ or $\sin(2\,\beta)$ given in 
%Equations (\ref{eq:eta1})-(\ref{eq:eta2}), is due to an optimistic use 
%of the theoretical uncertainties attached to $\fbdsqbd$.  
The present analysis shows that these results are in practice very weakly dependent 
on the  exact value taken for the uncertainty on $\fbdsqbd$.
An evaluation of this effect has been already presented in~\cite{ref:taipei}
where the flat part of the theoretical uncertainties on $\fbdsqbd$
was  multiplied by two. Similar tests will be shown in Section \ref{sec:stability}.

\subsection{Further comments on the Lattice predictions for \boldmath$\fbdsqbd$ and $\hat B_K$}
\label{sec:bkfbtheo}

The values found in the present analysis for $\fbdsqbd$ and $\hat B_K$ are in agreement with those
of previous studies~\cite{ref:parodietal}--\cite{ref:herab}.
They are also in agreement with  the  predictions  from lattice 
QCD.  It can be noticed that lattice predictions for these quantities existed well before it were possible to extract
them from the analysis of the unitarity triangle.
Moreover these predictions  have been stable over the years: one of the first calculations
of $\BK$ gave $\BK^{\overline{\rm MS}}(\mu=2 \, {\rm GeV})=0.65 \pm 0.15$~\cite{gavela}
in 1987 (corresponding to $\hat \BK= 0.90 \pm 0.20$)
and $\hat B_K$ was estimated to be  $\hat B_K = 0.84 \pm 0.03 \pm 0.14$ in 1996~\cite{sharpe96}; a compilation  
by one of the authors of the present paper gave   $f_{B_d}  \sqrt{\hat B_{B_d}}= (220 \pm 40) $ MeV 
and  $f_{B_d} \sqrt{\hat B_{B_d}}= (207 \pm 30)$ MeV,   in 1995 and 1996 
respectively~\cite{marti96}.

\subsection{Lower bounds on \boldmath$f_{B_d}\sqrt{\hat B_{B_d}}$ and \boldmath$\hat B_K$}
\begin{figure}[htb!]
\begin{center}
\epsfig{file=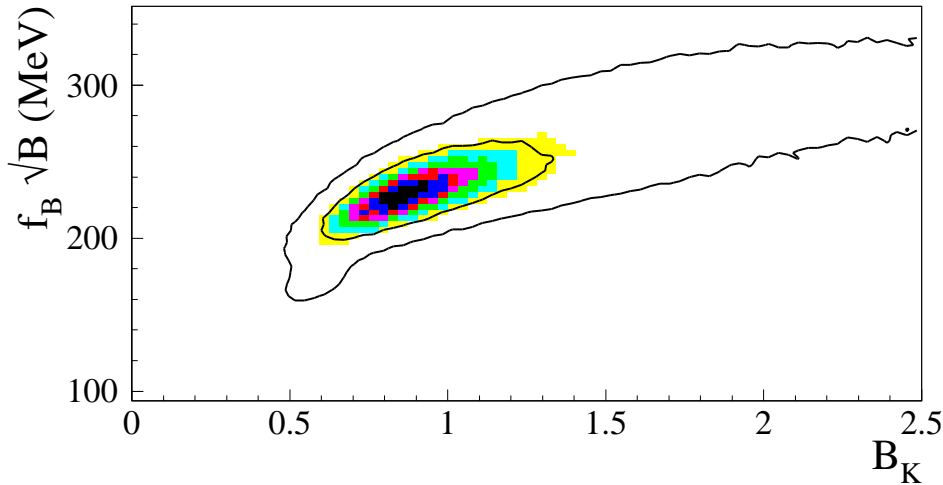,width=15cm}
\end{center}
\caption{\small{ The 68\% and 95\% contours in the $(f_{B_d}\sqrt{\hat B_{B_d}}$,$\hat B_K)$ plane.}} 
\label{fig:bkfb}
\end{figure}

The  region in the plane ($f_{B_d}\sqrt{\hat B_{B_d}}$,~$\hat B_K$), which is obtained
by removing the theoretical constraints on these quantities, is shown in Figure~\ref{fig:bkfb}.
It appears that present constraints can cope with very large values of $\hat B_K$ and thus it was needed
to restrict the possible range of variation for this parameter. In the present study, probabilities 
have been normalised assuming $\hat B_K<\rm{5}$.
The 68\% and 95\% probability contours have rather different shapes.
Within 68\% probability,  both $f_{B_d}\sqrt{\hat B_{B_d}}$ and $\hat B_K$  are well constrained.
The most important conclusion which can be drawn from this study is the simultaneous lower bounds on
$f_{B_d}\sqrt{\hat B_{B_d}}$ and $\hat B_K$, namely
\begin{equation}
\hat B_K > 0.5\, \quad {\rm and} \quad \quad f_{B_d}\sqrt{\hat B_{B_d}} >150\,\,\mbox{MeV}\,,
\quad {\rm at~ 95}\%~\mbox{probability}.
\end{equation}
%Whereas the lower bound on $f_{B_d}\sqrt{\hat B_{B_d}}$ is lowered by
%relaxing the $\Delta m_s$ constraint, the bound on $\hat B_K$ is practically unchanged,
%since it depends on $\epsilon_K$ and $\Vub$ only. 

%%%%%%%%%%%%%%%%%%%%%%%%%%%%%%%%%%%%%%%%%%%%
%%%%%%%%%%%%%%%%%%%%%%%%%%%%%%%%%%%%%%%%%%%%%%%

\subsection{Other quantities of interest}
For completeness the values of the phases $\chi$, $\chi^\prime$ and $\mbox{Im}\,\lambda_t$ 
\cite{ref:angol} are also given.

\begin{eqnarray}
\chi &=& {\rm arg}\left(-\frac{V_{cs}^* V_{cb}}{V_{ts}^* V_{tb}}\right) \quad
= {\rm atan}\left(\frac{\etabar}{1+\lambda^2 \rhobar}\right) \, , \nonumber \\
%&\,& \nonumber \\
\chi^\prime &=& {\rm arg}\left(-\frac{V_{ud}^* V_{us}}{V_{cd}^* V_{cs}}\right)
\quad = 
 {\rm atan}\left(\frac{(\Vcb^2 \etabar)}
 {1-\lambda^2/2 -(\Vcb^2/2)(1-2 \rhobar)}\right) \, , \nonumber \\
\quad \mbox{Im}\,\lambda_t \quad &=& \lambda \Vcb^2 \etabar \, .
\label{eq:chi} 
\end{eqnarray} 
The measured values of these parameters are
\begin{eqnarray}
\chi = (17.4 \pm 2.1)^{\circ} ,\quad & ~\chi^{\prime}=(0.031 \pm 0.003)^{\circ} ,\quad &
\mbox{Im}\,\lambda_t = (1.19 \pm 0.12)~ 10^{-4}  \, .
\label{eq:angoletti}
\end{eqnarray}

\section{Stability of the results}
\label{sec:stability}
The sensitivity of present results on the assumed probability distributions attached 
to the input parameters was studied.
The comparison of the results obtained by varying the size of the
theoretical  uncertainties
has been done to evaluate the sensitivity to these variations of uncertainties quoted
on fitted values. {\it This must not be taken as a proposal to inflate
the uncertainties obtained in the present analysis}.

\begin{table}[htb!]
\begin{center}
\footnotesize{
\begin{tabular}{|c|c|c|c|}
\hline

Parameter(s)  & Modified Value(s) & $\rhobar$ & $\etabar$ \\
\hline

     All                           &           None              & $0.224 \pm 0.038$ & $0.317 \pm 0.040$ \\
${\hat \BK}$                       & $0.87 \pm 0.06 \pm 0.26$    & $0.225 \pm 0.040$ & $0.314 \pm 0.045$ \\
$\fbdsqbd$                         & $(230 \pm 25 \pm 40)\MeV$   & $0.224 \pm 0.039$ & $0.318 \pm 0.041$ \\
$\xi$                              & 1.14 $\pm$ 0.04 $\pm$ 0.10  & $0.221 \pm 0.045$ & $0.318 \pm 0.042$ \\
$\Vub$                             & (37.8 $\pm$ 4.9) 10$^{-4}$  & $0.240 \pm 0.044$ & $0.334 \pm 0.048$ \\
$\Vcb$                             & (40.7 $\pm$ 2.6) 10$^{-3}$  & $0.224 \pm 0.039$ & $0.320 \pm 0.048$ \\
$\dms$ cons.                       & see Sect.~\ref{sec:deltams} & $0.227 \pm 0.047$ & $0.315 \pm 0.042$ \\
All                                &    as above                 & $0.249 \pm 0.064$ & $0.333 \pm 0.065$ \\
\hline
\end{tabular}
\caption[]{ \small { Stability tests (1). Central values and uncertainties for $\rhobar$ and $\etabar$ 
have been defined relative to the median of the corresponding p.d.f. }}
\label{tab:stab1} }
\end{center}
\end{table}

\begin{table}[htb!]
\begin{center}
\footnotesize{
\begin{tabular}{|c|c|c|c|c|}
\hline

Parameter(s)  & Modified value(s) & $\sin{(2\beta)}$ & $\sin{(2\alpha)}$ & $\gamma$(degrees) \\
\hline
      All     &          None                   & $0.698 \pm 0.066$          & $-0.42 \pm 0.23$ & $54.8 \pm 6.2$ \\
${\hat \BK}$  & $0.87 \pm 0.06 \pm 0.26$        & $0.693 \pm 0.071$          & $-0.45 \pm 0.26$ & $54.5 \pm 6.7$ \\
$\fbdsqbd$    & $(230 \pm 25 \pm 40)\MeV$       & $0.698 \pm 0.067$          & $-0.44 \pm 0.24$ & $54.9 \pm 6.3$ \\
$\xi$         & 1.14 $\pm$ 0.04 $\pm$ 0.10      & $0.697 \pm 0.067$          & $-0.42 \pm 0.26$ & $55.4 \pm 7.2$ \\
$\Vub$        & (37.8 $\pm$ 4.9) 10$^{-4}$      & $0.733 \pm 0.080$          & $-0.40 \pm 0.24$ & $54.4 \pm 6.1$ \\
$\Vcb$        & (40.7 $\pm$ 2.6) 10$^{-3}$      & $0.701 \pm 0.077$          & $-0.43 \pm 0.26$ & $54.8 \pm 6.6$ \\
$\dms$ cons.  & see Sect.~\ref{sec:deltams}     & $0.698 \pm 0.067$          & $-0.45 \pm 0.27$ & $54.6 \pm 7.6$ \\
All           &         as above                 & $0.738^{+0.086}_{-0.101}$ & $-0.42 \pm 0.37$ & $53.3 \pm 10.0$ \\
\hline
\end{tabular}
\caption[]{ \small { Stability tests (2). Central values and uncertainties for
$\sin{(2\beta)}$, $\sin{(2\alpha)}$ and $\gamma$ have been defined relative to the median of the corresponding p.d.f. }}
\label{tab:stab2} }
\end{center}
\end{table}

In these tests, all values for uncertainties of theoretical origin have been, in turn,
multiplied by two. For the quantities $\Vub$ and $\Vcb$, new p.d.f. have been
determined, following the prescriptions given in Sections \ref{secvcb} and \ref{secvub} 
and used in the analysis; central values and uncertainties
quoted in Tables \ref{tab:stab1} and \ref{tab:stab2} correspond to the average and standard deviation
of these distributions.

The main conclusion of this exercise is that, even in the case where all theoretical uncertainties are doubled,
the unitarity triangle parameters are determined with an uncertainty which increases only by about 1.5 
(see Tables \ref{tab:stab1} and \ref{tab:stab2}).

\section{Comparison between the {\bf standard} and the {\bf  95\%\,C.L. scanning}
approaches}
\label{sec:comparison}

\begin{table}
\begin{center}
\begin{tabular}{|c|c|}
\hline
                       parameter                 &  Value  $\pm$ Gaussian
$\pm$ Flat errors\\
\hline
$\left | V_{cb} \right |$  & $(40.0 \pm 2.0 \pm 0.0)\times 10^{-3}$ \\
$\left | V_{ub} \right |$ & $(34.0 \pm 1.4 \pm 5.0)\times 10^{-4}$ \\
$\Delta m_d$              & $(0.473 \pm 0.016 \pm 0.000)$ ~ps$^{-1}$\\
$\Delta m_s$          & $>$ 14.3 ps$^{-1}$  \\
$\epsilonk$                        &
$(2.285\pm 0.018 \pm 0.000) \times 10^{-3}$ \\
$\BK$                & $0.80 \pm 0.00 \pm 0.15$ \\
$ f_{B_d} \sqrt{\hat B_{B_d}}$      & $(200 \pm 0 \pm 40)~\MeV$\\
$\xi=\frac{ f_{B_s}\sqrt{\hat B_{B_s}}}{ f_{B_d}\sqrt{\hat B_{B_d}}}$ &
$1.14 \pm 0.00 \pm 0.08$ \\
\hline
\end{tabular}
\caption[]{ {Values of the quantities used in the comparison between
the standard and the {\it 95\%\,C.L. scanning} methods. The quoted flat error values are used
to define the range of variation of the corresponding
parameter in the {\it 95\%\,C.L. scanning} approach
and to define the uniform probability distribution for this parameter in the
standard
method.}}
\label{tab:2}
\end{center}
\end{table}

The theoretical basis which allows to define, in the standard approach, regions
of the $(\rhobar, \etabar)$ plane  which correspond to any
given value of confidence has been already discussed.
For completeness, in this Section, a comparison of present
results at the 95\% C.L. with the corresponding regions selected using the
{\it 95\%\,C.L. scanning} approach is made.  In the latter approach, this region 
is the envelope of 95\% C.L. contours defined at several points  and, as discussed previously,
it is not easy to understand  to which level of confidence they correspond to.

The results of the study given in~\cite{ref:plasz} have been used in this comparison and the
same central values for the parameters have been used in the two cases.
When, in the  {\it 95\%\,C.L. scanning} approach,  a parameter is scanned over
a given interval,  in the  standard approach a flat probability distribution defined 
over the same range has been used. In Table \ref{tab:2} central values and uncertainties used in this comparison,
taken from~\cite{ref:plasz}, have been collected.

%%%%%%%%%%%%%%%%%%%%%%%%%%%%%%%%%%%%%%%%%
\begin{table}[htb!]
\begin{center}
\begin{tabular}{|c|c|c|}
\hline
                       parameter                 & standard approach &
 {\it 95\%\,C.L. scanning} approach\\
                       & (95\%\, prob. range) &    \\
\hline
$\rhobar$  & $-0.06,~0.31$ & $0.00,~0.30$ \\
$\etabar$ & $0.26, ~0.42$ & $0.20,~0.45$\\
$\sin(2\,\beta)$ & $ 0.56, ~0.82$ & $0.50,~0.85$\\
\hline
\end{tabular}
\caption[]{\small {
Comparison of standard and {\it 95\%\,C.L. scanning} approaches.
Note that intervals quoted for the results of the standard approach
correspond to 95\% probability for each quantity, whereas in the
{\it 95\%\,C.L. scanning} method the intervals are obtained from  the maximum and minimum values
spanned by the 2-D envelope.
}}
\label{tab:rhoeta} 
\end{center}
\end{table} 

%%%%%%%%%%%%%%%%%%%%%%%%%%%%%%%%%%%%%%%%%%%%
\begin{figure}[htb!]
\begin{center}
\epsfig{file=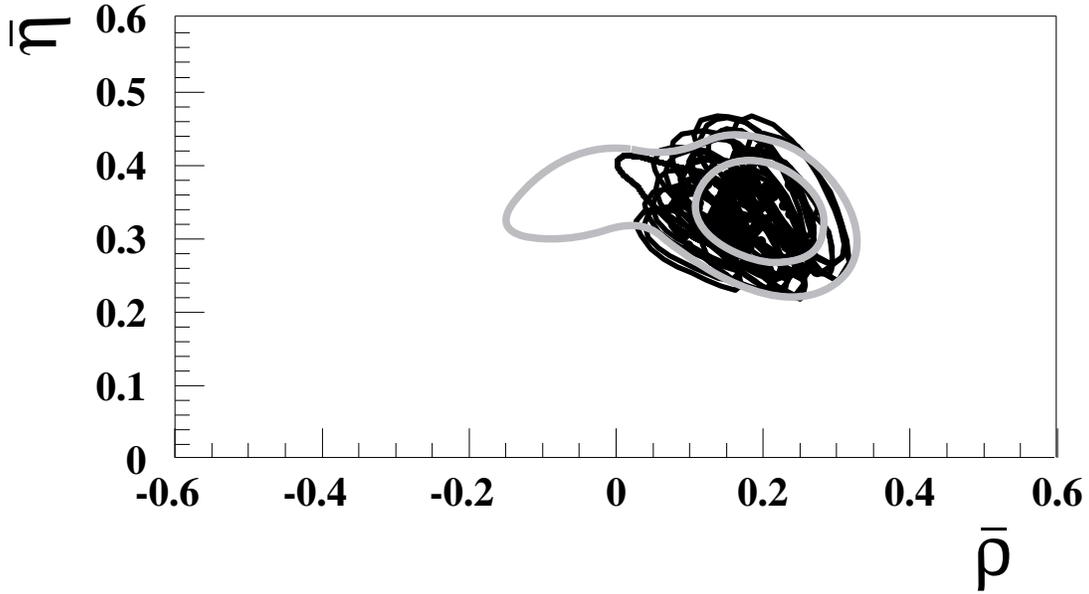,width=15cm}
\end{center}
\caption{\small { The contour corresponding to 95\% probability in the standard
approach, given by the external continuous line, has been compared with the region spanned by  
95\% C.L. ellipses obtained in the {\it 95\%\,C.L. scanning} method. The internal contour, 
corresponding to 68\% probability in the standard approach, is also drawn.}} 
\label{fig:scanning standard}
\end{figure}
%%%%%%%%%%%%%%%%%%%%%%%%%%%%%%%%%%%%%%%%%%%%%%%
\begin{figure}[htb!]
\begin{center}
\begin{tabular}{cc}
\epsfig{file=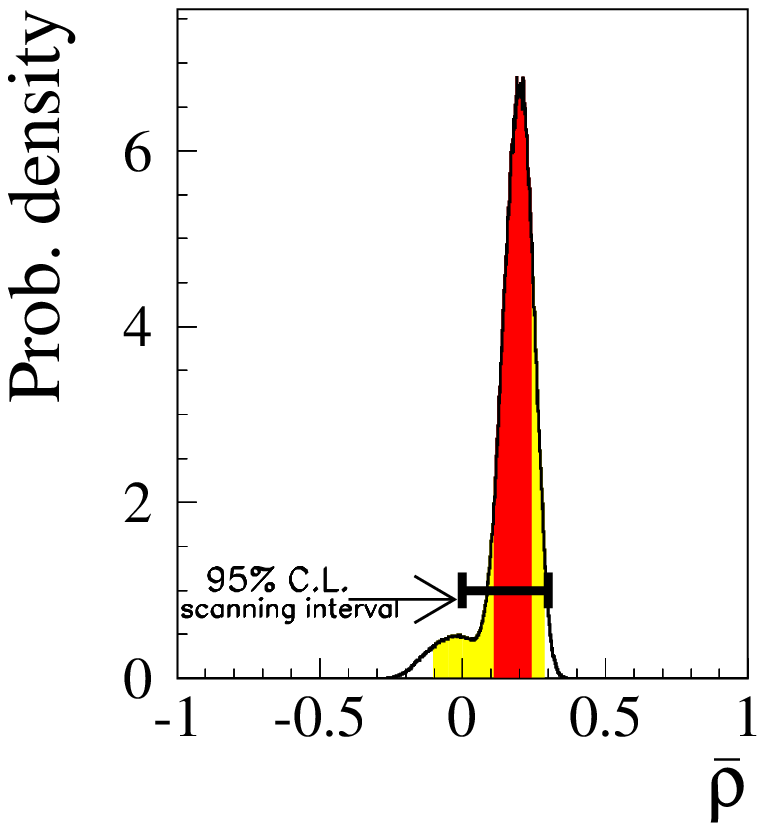,bbllx=30pt,bburx=270pt,bblly=1pt,bbury=240pt,width=8cm} & 
\epsfig{figure=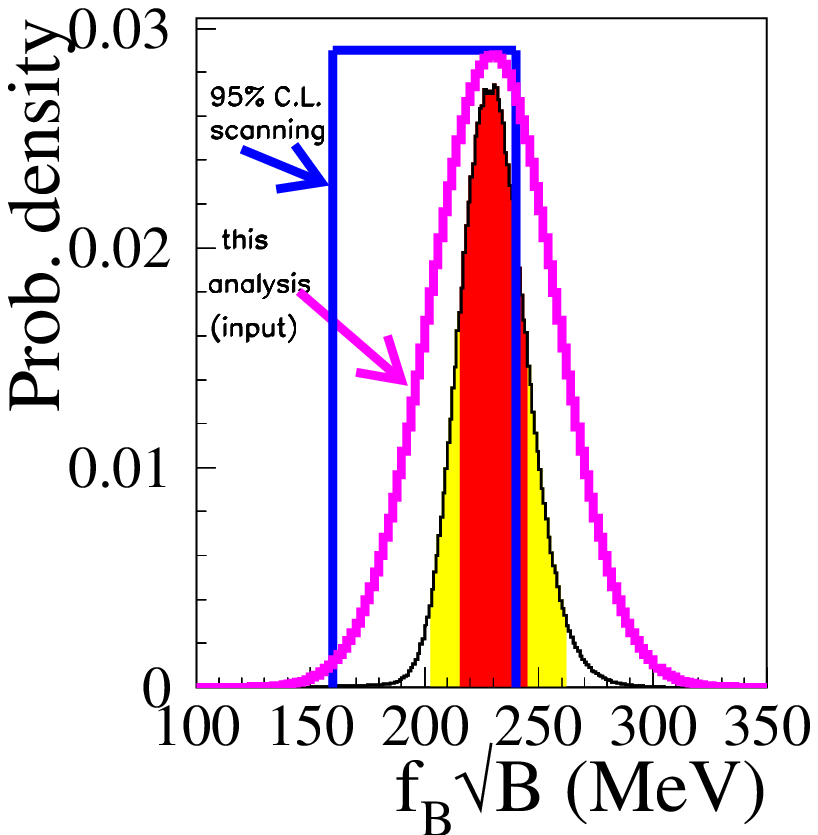,bbllx=40pt,bburx=300pt,bblly=1pt,bbury=240pt,height=8cm}
\end{tabular}
\end{center}
\caption{\small{ Left: Probability distribution of the $\rhobar$ variable obtained in the
standard approach. The interval selected in the `{\it 95\%\,C.L. scanning} method
has been indicated. Right: Comparison between different input distributions for the parameter
$ f_{B_d} \sqrt{\hat B_{B_d}}$ and the final probability distribution
obtained with all constraints but the measurement of $\dmd$.}} 
\label{fig:scanbis}
\end{figure}
%%%%%%%%%%%%%%%%%%%%%%%%%%%%%%%%%%%%%%%%%%%%%%%%%%%%%

The constraint coming from the study of ${B}^0_s-\bar{{B}}^0_s$
oscillations is included using the weight given in Equation~(\ref{dms_wrong}), since
it was used also in the {\it 95\%\,C.L. scanning} analysis.

In Figure~\ref{fig:scanning standard} the ``95\% C.L.'' contours obtained with the two methods have
been compared and the 95\% C.L. intervals for $\rhobar$ and $\etabar$
are given in Table \ref{tab:rhoeta}.
In Figure~\ref{fig:scanbis}-left the distribution of $\rhobar$ is also shown.

Several remarks are in order:
\begin{itemize}
\item when using the same values for the input parameters and for their 
uncertainties, the {\it 95\%\,C.L. scanning} and the standard approaches 
select similar regions in the $(\rhobar,~\etabar)$ plane. 
It is then unjustified to qualify as too optimistic the standard approach. 
The main difference is precisely the opposite, namely that,
unlike the case of {\it 95\%\,C.L. scanning}, with the standard
method one is able to {\it quantify} the confidence corresponding to 
any given interval. 
\item
in the standard approach the probability distribution in the 
$(\rhobar,~\etabar$) plane is meaningful. In the present exercise, the interval
$[0.15,~0.30]$ for $\rhobar$ is more probable than $[0.,~0.15]$.
Quoting simply that the value of $\rhobar$ is expected to be between
0 and 0.3, at 95\% C.L. does not indicate that there is a most
probable value around 0.2.
Such considerations are not  in order in the  {\it 95\%\,C.L. scanning} approach;
\item
even in the {\it 95\%\,C.L. scanning} approach it is necessary to define intervals
which are expected to contain the true values of the parameters. But then
some probability has to be attributed to these intervals. It is thus 
illusory the statement that with this method one does not attribute
any probability distribution to the theoretical inputs.
Indeed the statement that an input cannot assume a value outside a given 
interval, is equivalent to attribute zero probability 
to the region outside the interval; 
\item  
the flat distribution for a theoretical 
parameter, in some cases,   is a very strong a-priori, as illustrated 
in Figure~\ref{fig:scanbis}-right, where 
 the choice of the central value and  error on $f_B \sqrt{\hat B_B}$ used 
in~\cite{ref:plasz} is shown.
The flat interval
corresponds to the values quoted in Table~\ref{tab:2}. 
The broad Gaussian-like distribution, used in the present analysis,  results
from a convolution between a flat and a Gaussian distributions using the
parameters of Table~\ref{tab:1}.  This is also shown in the figure,
together with the final p.d.f. resulting from our analysis.
From this comparison,  it appears that the region favoured by present measurements is
marginally compatible with the interval selected in the {\it 95\%\,C.L. scanning}.
In particular, the latter 
excludes {\it a priori} (large) values of $f_B \sqrt{\hat B_B}$ which are
still compatible with data.
\end{itemize}

\section{Conclusions}
\label{sec:conclusions} 
 In this paper, we  have discussed the inferential framework which allows 
a consistent analysis of the unitarity triangle,
where both experimental and theoretical uncertainties 
play an important role. In particular, we were concerned with the treatment 
of the {\it theoretical errors}, which raised several 
controversies in the past. We have shown that, in this framework, 
there is a consistent way of handling the theoretical 
uncertainties and that  these  can
be included in the analysis in a way similar to the experimental ones.   
The main results of our analysis have been discussed  in
Section~\ref{sec:results}  and, for convenience, are listed below (see Table \ref{tab:finalresults}).
\par  Present measurements of $\epsilonk$, $\vubsvcb$, $\dmd$ and
of the limit on $\dms$ allow, within the framework of the Standard Model,
and after including results from lattice QCD on $\hat \BK$,
$\fbdsqbd$ and $\xi$, to determine the parameters of the CKM unitarity triangle.

\begin{table}[htb!]
\begin{center}
\begin{tabular}{|c|c|}
\hline
 Parameters  &  Present analysis \\
\hline

$\rhobar$         &  0.224 $\pm$ 0.038         \\
$\etabar$         &  0.317 $\pm$ 0.040         \\
$\sin(2\,\beta)$  &  0.698 $\pm$ 0.066         \\
$\sin(2\alpha)$   &  -0.42 $\pm$  0.23         \\
$\gamma$          & (54.8 $\pm$ 6.2)$^{\circ}$ \\
\hline 
\end{tabular}
\caption[]{ \small {The results on the unitarity triangle parameters presented in this paper.}}
\label{tab:finalresults} 
\end{center}
\end{table} 

The dependence  of  central values and  quoted uncertainties on 
 values and uncertainties assumed for the theoretical
parameters, and on  different procedures  to include the available information,
have been evaluated.
These exercises demonstrate that the standard approach employed to
 get our  numbers 
is on a sound basis and that results and uncertainties for the
quantities of interest are stable.
The determination of central values and uncertainty distributions for 
the theoretical inputs  used in the present analysis
has been reviewed and explained.

By removing, in turn, the different  constraints it has been verified that,
within the present uncertainties, the Standard Model description
of CP violation is in agreement with the measurements. In particular, $B$ oscillations 
and decays select, in the $(\rhobar,~\etabar)$ plane, a similar region
as $\epsilonk$  for the $K^0$ system.
This constitutes already a test of compatibility between the measurements of the sides  
and of the angles of the CKM triangle. Its accuracy is given by the quoted uncertainties on $\hat \BK$
appearing in the first row of Table \ref{tab:sum}.\\
This Table gives a summary of the determination of the values of some theoretical parameters
and of $\dms$ obtained in the present analysis and corresponding estimates from lattice QCD:

\begin{table}[htb!]
\begin{center}
\begin{tabular}{|c|c|c|c|}
\hline
 Parameters &      Present analysis      &       Present analysis         &  Lattice QCD  \\ 
            &  (with the contraint)      &  (without the contraint)       &               \\ 
\hline

$\hat \BK$ & $0.88 \pm 0.08$           & $0.90^{+0.30}_{-0.14}$   &  $0.87 \pm 0.06 \pm 0.13$ \\

$\fbdsqbd$ & $(230 \pm 12)$ MeV        & $(231 \pm 15)$ MeV       &  $(230 \pm 25 \pm 20)$ MeV \\

$\dms$ &  $(17.3^{+1.5}_{-0.7})$~ps$^{-1}$ & $(16.3 \pm 3.4)$~ps$^{-1}$ & $(15.8 \pm 2.3 \pm 3.3)$~ps$^{-1}$\\

\hline 
\end{tabular}
\caption[]{ \small { Comparison between the values of the parameters determined in the present
analysis with those obtained by lattice QCD. In the second row (third row) the results are obtained 
including (not including) the given constraint.}}
\label{tab:sum} 
\end{center}
\end{table} 

Values of the parameters given in Table \ref{tab:sum} and of the angles reported in Table \ref{tab:finalresults} 
can be considered as reference numbers to which new measurements, using different approaches, can be 
compared in view of identifying new physics contributions. 

Three results will become available in a near future. 
\begin{itemize}
\item
Analyses of $B$-meson two body decays have already
produced values for $\gamma$ which were more than 2$\sigma$ larger than the value given in 
Table \ref{tab:finalresults}.
Including recent results form BaBar and Belle
     Collaborations, smaller values of the angle $\gamma$ are
     preferred~\cite{ref:babarsin2b},\cite{ref:bellesin2b}. Before claiming evidence of new physics, a      
     better theoretical understanding of non-leptonic decays and more stable  
     measurements are needed.
\item
Direct measurements of sin2$\beta$ obtained studying $J/\psi K_S$ events at $B$ factories or at TeVatron 
will provide important tests when the accuracy of these measurements will be better than 0.1. 
\item
Finally, a first evidence of ${B}^0_s-\bar{{B }}^0_s$ oscillations could still come from LEP and SLD Collaborations and 
accurate measurements are expected from the TeVatron. Measured values larger than 25~$\rm {ps^{-1}}$ could provide 
evidence for new physics.
\end{itemize}

\section*{Acknowledgements}
We warmly thank D.~Becirevic, C.~Bernard, V.~Gimenez  and S.~Sharpe for very useful
discussions about lattice hadronic parameters used in this study and their errors. 
We would like to warmly thanks A.~Buras for the careful reading of this paper.
G.M. thanks LAL, LPT and \'Ecole Polytechnique, where part of this work was done, for
the kind hospitality. V.L. and G.M. acknowledge MURST for partial support.   
M.C. and E.F. thank the T31 group for the kind hospitality at the TU M\"unchen, 
where part of this work was done.

\end{document}